# Large enhancement of near-field radiative heat transfer in the dual nanoscale regime enabled by electromagnetic corner and edge modes

Lei Tang[1], Lívia M. Corrêa[2], Mathieu Francoeur[3,*], and Chris Dames[1,*]

[1]Department of Mechanical Engineering, University of California at Berkeley, Berkeley, CA, USA
[2]Department of Mechanical Engineering, University of Utah, Salt Lake City, UT, USA
[3]Department of Mechanical Engineering, McGill University, Montréal, QC, Canada
[*]Corresponding authors: mathieu.francoeur@mcgill.ca, cdames@berkeley.edu

**Abstract**

It is well established that near-field radiative heat transfer (NFRHT) can exceed Planck's blackbody limit[1] by orders of magnitude owing to the tunneling of evanescent electromagnetic frustrated and surface modes[2-4], as has been demonstrated experimentally for NFRHT between two large parallel surfaces[5-7] and between two subwavelength membranes[8,9]. However, while nanostructures can also sustain a much richer variety of localized electromagnetic modes at their corners and edges,[10,11] the contributions of such additional modes to further enhancing NFRHT remain unexplored. Here, for the first time, we demonstrate both theoretically and experimentally a new physical mechanism of NFRHT mediated by these corner and edge modes, and show it can dominate the NFRHT in the "dual nanoscale regime" in which both the thickness of the emitter and receiver, and their gap spacing, are much smaller than the thermal photon wavelengths. For two coplanar 20 nm thick SiC membranes separated by a 100 nm vacuum gap, the NFRHT coefficient at room temperature is both predicted and measured to be 830 W/m²K, which is 5.5 times larger than that for two infinite SiC surfaces separated by the same gap, and 1400 times larger than the corresponding blackbody limit accounting for the geometric view factor between the emitter and receiver. This enhancement is dominated by the electromagnetic corner and edge modes which account for 81% of the NFRHT between these SiC membranes. These findings are important for future NFRHT applications in thermal management and energy conversion.

The NFRHT between micrometer-[5,6,12,13] and millimeter-scale[7,14-18] parallel flat surfaces and between two subwavelength membranes[8,9] separated by a nanoscale vacuum gap $d$ has been well established to exceed the blackbody limit[1], in agreement with fluctuational electrodynamics predictions[2,3,19,20]. The mechanism responsible for this enhancement is well understood and is mediated by the tunneling of evanescent electromagnetic waves, which include frustrated and surface modes[2-4]. Such near-field enhancements to the heat transfer are expected to become important only when $d$ is much smaller than the characteristic thermal wavelengths of the corresponding free-space photons, $\lambda_{th}$.

The physics of NFRHT becomes even richer when the size of the emitter and receiver themselves, for example as characterized by their transverse breadth (membrane thickness $t$ in Fig. 1), is also in the deep subwavelength regime. In this case, the membranes can also sustain localized electromagnetic corner and edge modes which are otherwise forbidden by the translational symmetry of infinite parallel surfaces and play only a negligible role in NFRHT between relatively thick membranes[8,9]. Such modes have previously been studied in non-thermal contexts like plasmonic phenomena in nanoscale metal structures.[10,11] Unexplored, however, is the potential for such corner and edge modes to couple across two nanostructures in the near field and thereby contribute to NFRHT.

For the first time, here we investigate both theoretically and experimentally the NFRHT in this "dual nanoscale" regime, in which both $d$ and $t << \lambda_{th}$. As depicted in Fig. 1a, we study the NFRHT between several pairs of coplanar SiC membranes with $t$ and $d$ both much smaller than the relevant $\lambda_{th} \gtrsim 7$ μm. We find that the new physical mechanism introduced by electromagnetic corner and edge modes leads to large enhancements of NFRHT compared to the case of two infinite parallel surfaces, and indeed can dominate the NFRHT for $t \ll d \ll \lambda_{\text{th}}$. For 20 nm thick SiC membranes separated by a vacuum gap of 100 nm, the room temperature radiative heat transfer coefficient $h_{\text{rad}}$ = 830 W/m$^2$K, with excellent agreement between theory and measurement. This $h_{\text{rad}}$ is very large, being 5.5 times larger than calculated for two infinite SiC surfaces separated by the same gap, and 1400 times larger than the blackbody limit accounting for the geometric view factor between the emitter and receiver face areas. This large NFRHT in the dual nanoscale regime might be exploited in a variety of applications, such as localized radiative cooling[21], thermal management technologies[22-26], and energy conversion devices[27-31].

**Devices for measuring NFRHT between coplanar SiC membranes**

We measure NFRHT in the dual nanoscale regime using three main microfabricated devices consisting of pairs of suspended coplanar SiC membranes with thicknesses of 20, 50, and 120 nm (Fig. 1b). The choice of SiC membrane dimensions was guided by preliminary numerical simulations of NFRHT using the discrete system Green's function (DSGF) method.[32]

Each device consists of two symmetrical islands, the heating island at temperature $T_{\text{h}}$ and the cold island at $T_{\text{c}}$. Each island is supported by six long suspended beams (Pt on SiNx) and incorporates a serpentine Pt resistance thermometer (PRT) for heating and thermometry. The NFRHT is between the thin symmetrical SiC membranes separated by a nanosized gap $d$. The width $w$ of all devices is 30 μm, and the suspended portions of the SiC membranes have lengths $L = 9$ μm for the 120 nm thick device and $L = 1$ μm for the 50 and 20 nm thick devices. Another $t = 20$ nm, $w = 30$ μm device was prepared with $L = 2$ μm.

Two symmetric SiC "wings" are located on each island on the side opposite to the SiC membranes that carry the NFRHT. These wings have the exact same dimensions as the inner SiC membranes, and are included to increase the yield of the devices during release. The SiNx of the islands is patterned to slightly overlap the contact edges of the SiC membranes and wings by ~1 μm to minimize the thermal contact resistance. More details about the microfabrication are given in Supplementary Section 1, and about the device design in our recent work measuring the thermal conductivity of similar SiC films.[33]

Figure 1(c,d) respectively show optical and scanning electron microscope (SEM) images of the 120 nm thick device. The gap spacing $d$ is estimated from tilted SEM images of the membrane cross-sections prior to the final release step (the KOH release etch of Si has negligible impact on the SiC, see Supplementary Section 1). For the two thicker samples, Fig. 1e shows how the SiC etch resulted in sidewalls that are sloped rather than being perfectly vertical. For example, for the 120 nm thick device the gap spacing ranges from 85 nm to 245 nm, with an average gap spacing $\bar{d} = 165$ nm. The DSGF simulations of NFRHT below use this actual sloped gap geometry. Similarly, the 50 nm thick device has a gap sloping from 110 to 160 nm ($\bar{d} = 135$ nm), while both 20 nm thick devices have an approximately uniform gap of $d = \bar{d} = 100$ nm.

The flatness and coplanarity of the suspended SiC membranes after their release is another key consideration, which we characterize using 3D laser confocal microscopy (LCM). For all devices the LCM topography scans confirmed very good alignment of the suspended SiC membrane edges across the gap, with a misalignment of no more than ±20 nm when comparing the $z$-heights of the left and right SiC membranes' free edges at the gap (see Supplementary Table 1 and Supplementary Fig. 2). Similarly, LCM was also used to quantify the slight tilt of each suspended $L \times w$ membrane away from its ideal $x$-$y$ plane; as expected these tilts were small, never more than 5°, and also quite symmetric about the gap (see Supplementary Sections 2 and 3). Additional DSGF simulations of such residual $z$-misalignment and tilt effects indicate they should have negligible impact on $h_{rad}$ (less than 12%; see Methods).

**a**

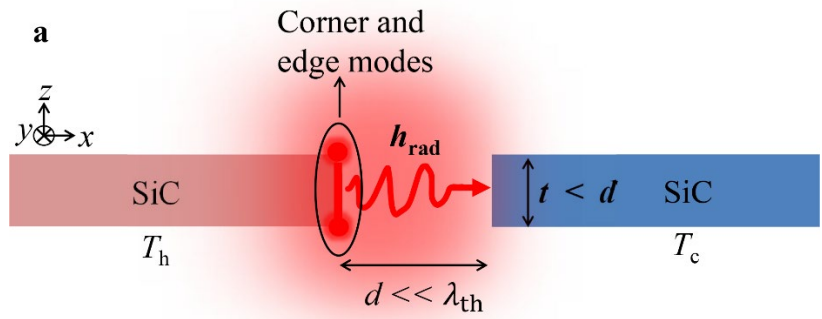

**b**

**Top View**

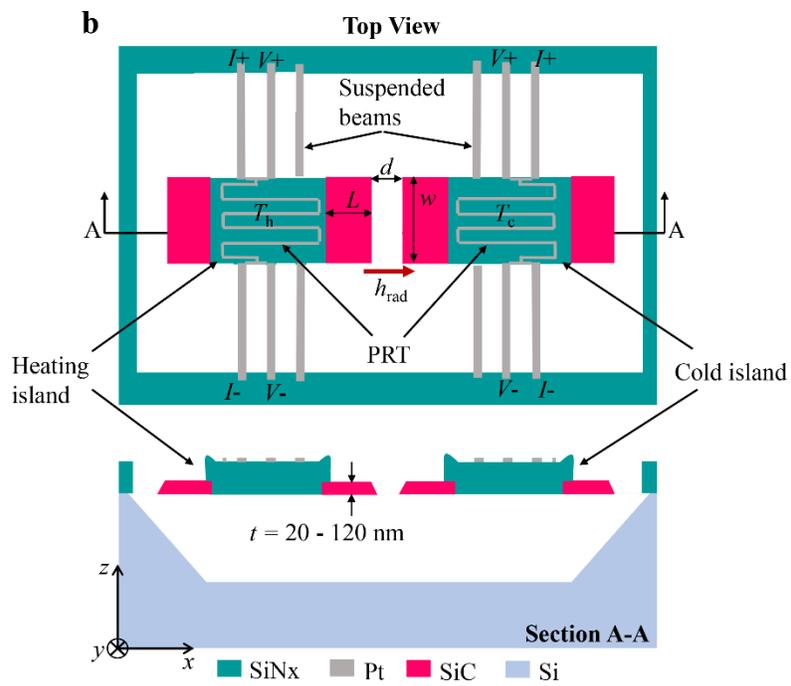

Section A-A

**c**

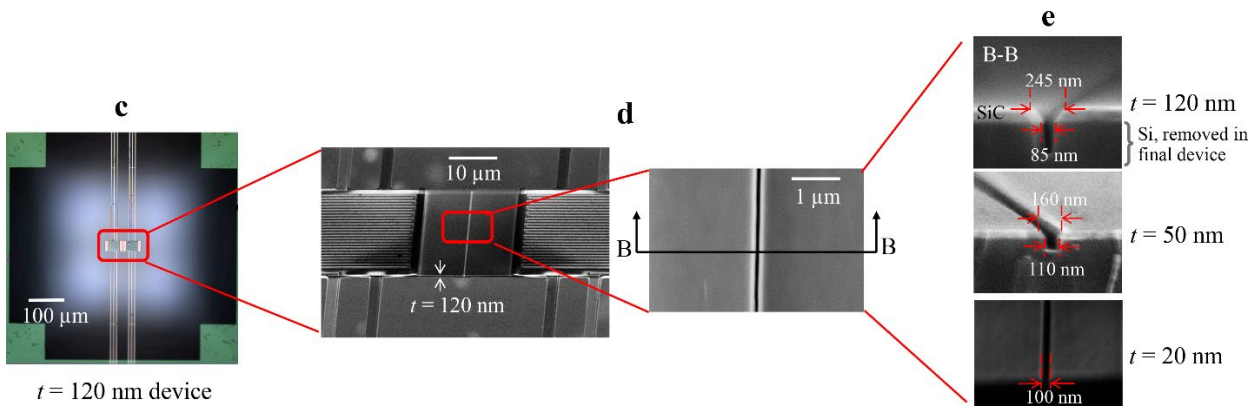

*t* = 120 nm device

**d**

**e**

**Fig. 1 | Principle of NFRHT in the dual nanoscale regime and a device for measuring it. a,** Concept of NFRHT between two thin coplanar SiC membranes mediated by electromagnetic corner and edge modes. **b,** Schematic of the measurement device. Matched SiC membranes separated by a gap $d$ are supported by symmetric SiNx islands, each with a Pt serpentine for Joule heating and resistance thermometry to measure the NFRHT. The three main devices studied in this work use paired SiC membranes with thicknesses $t$ = 20, 50, and 120 nm, and average gap spacings of $\bar{d}$ =100, 135, and 165 nm, respectively. **c,** Optical and **d,** SEM images of the released 120 nm thick device. **e,** SEM images of the SiC gap details of the three main devices, after the etch to define the gap but prior to removing the sacrificial Si underlayer. The 20 nm thick device has an approximately uniform gap $\bar{d}$ = 100 nm, while the 120 and 50 nm thick devices exhibit sloped gaps with $d$ from 85 nm to 245 nm ($\bar{d}$ = 165 nm) and from 110 nm to 160 nm ($\bar{d}$ = 135 nm), respectively. Such sloped $d$ is taken into account in the DSGF simulations of Fig. 2 (see also Supplementary Fig. 7).

## NFRHT measurement procedure

The NFRHT measurements follow a well-established electrothermal approach[33-37] (see also Supplementary Section 4), in which Joule heating at the hot island creates a small temperature difference across the gap that is measured using resistance thermometry of each island's PRT. AC lock-in methods are used to enhance accuracy, with typical heating current $I_f$ of amplitude 8 μA and frequency $f$ of 0.98 Hz. The resulting temperature rises of both islands at $2f$ are denoted $\Delta T_{h,2f}$ and $\Delta T_{c,2f}$. The typical temperature difference across the gap is $T_h - T_c \sim 1$ K, and measurements are repeated for cryostat stage temperature $T$ ranging from 200 K to 400 K. The cryostat operates at a vacuum level better than $5 \times 10^{-6}$ torr, such that convection and conduction through residual air are negligible. Finally, the total heat transfer coefficient from the heating island to the cold island is calculated as $h = \frac{G}{wt}$ where $G$ is the total thermal conductance. For this AC scheme Refs. [34,35] have shown that $G = \frac{\partial Q}{\partial(\Delta T_{h,2f} + \Delta T_{c,2f})} \cdot \frac{\partial \Delta T_{c,2f}}{\partial(\Delta T_{h,2f} - \Delta T_{c,2f})}$ , where $Q$ is the total Joule heating at the heating island's PRT plus one current-carrying Pt lead.

While the experimentally obtained $h$ between hot and cold islands is dominated by the NFRHT across the vacuum gap separating the SiC membranes, this total $h$ is also increased by contributions from any residual parasitic pathways. As detailed in Supplementary Section 5, the corresponding parasitic background conductance $G_{parasitic}$ was determined experimentally using a series of matched reference devices, and found to be never more than 35% of the total $G$, so that in all cases the raw measured $G$ of the test devices is dominated by the NFRHT contribution. Accordingly, the parasitic heat transfers have already been subtracted for all $h_{rad}$ results presented below in Fig. 2, a correction of no more than 35%.

**Experimental results and comparison with theory**

The measured $h_{rad}$ for the three main devices is shown by the solid-colored points in Fig. 2a with the experimental schematics shown in Fig. 2b, and Fig. 2c is a "slice" through those same results at $T = 300$ K to better highlight the trends with membrane thickness. In both panels of Fig. 2a and 2c, the hollow square point corresponds to the fourth device also with $t = 20$ nm and $d = 100$ nm, which was measured only at 300 K as an additional verification. The two 20 nm thick devices are nominally identical except for having two different lengths of the suspended SiC membrane: $L = 1$ μm for the primary device (solid red points in Fig. 2a) and $L = 2$ μm for the alternate device (hollow red point). Even with their different values of $L$ (and different $G_{parasitic}$), both 20 nm thick devices give nearly the same final value of $h_{rad}$, within 4%.

The color-coded shaded bands in Fig. 2 depict the corresponding DSGF simulations of $h_{rad}$. There is very good agreement between theory and experiment across all devices and temperatures. DSGF is a powerful and rigorous numerical approach to calculate the radiation heat transfer between 3D thermal emitters and receivers of arbitrary shapes and sizes based on discretization into cubic subvolumes[32] (see Methods for details). These DSGF simulations have no free parameters, with the only inputs being the temperature, the frequency-dependent dielectric function of SiC (see Methods), and the geometries of the emitter and receiver membranes. For all devices the finite breadth of the shaded bands of the DSGF predictions reflect the uncertainties arising from the small but finite tilting of the membranes as well as the actual nonuniform gap spacings (sloped $d$) of the 50 and 120 nm thick devices, as extracted from the LCM measurements and SEM images, respectively (see Methods). More details about the DSGF calculations and convergence are given in Supplementary Section 6.

A remarkable feature of the SiC NFRHT results in Fig. 2 is that $h_{rad}$ increases strongly with decreasing $t$. For further context, Fig. 2 includes theoretical results for two reference conditions. Reference Case 1 is a standard calculation (see Methods) for NFRHT between two infinite SiC surfaces ($t \rightarrow \infty$) separated by a gap of $d = 100$ nm, chosen to match the smallest $d$ of the experimental devices. At room temperature the measured $h_{rad}$ for the 20 nm thick devices reaches a very large value of 830 W/m²K (average of both 20 nm thick devices), which is 5.5 times larger than the $h_{rad}$ of 150 W/m²K predicted between two infinite surfaces. This 5.5-fold enhancement strikingly demonstrates the impact of additional physics of enhanced NFRHT in the "dual nanoscale" regime of small $d$ and $t$, as compared to the traditional "single nanoscale" regime of small $d$ alone[5-7,12-18] and the case of thick membranes.[8,9]

We further find that the SiC NFRHT in this dual nanoscale regime is dominated by evanescent waves rather than propagating waves (see Supplementary Section 7). As such, the physical mechanism underlying the heat transfer enhancement displayed in Fig. 2 is entirely different than that reported in Ref. 38, where radiative heat transfer between the subwavelength membranes was in the far-field regime and thus solely mediated by propagating waves.

Reference Case 2 is the blackbody limit in which all near-field and nanoscale effects are neglected. The two islands are treated as blackbodies, configured as parallel aligned rectangles each of area $wt$. This leads to $h_{BB} = 4\sigma_{SB}T^3F_{12}$, where $\sigma_{SB}$ is the Stefan-Boltzmann constant and $F_{12}$ is the geometric view factor between the two rectangles[39] (see Methods). Figure 2 shows that the measured NFRHT between SiC membranes in the dual nanoscale regime is vastly larger than the blackbody limit for the same geometry. For example, for 20 nm thick membranes separated by $d$ = 100 nm, the measured $h_{rad}$ at room temperature is around 1,400 times larger than the value of 0.61 W/m$^2$K for Reference Case 2. Similarly large enhancements are found for other temperatures (see Supplementary Section 8).

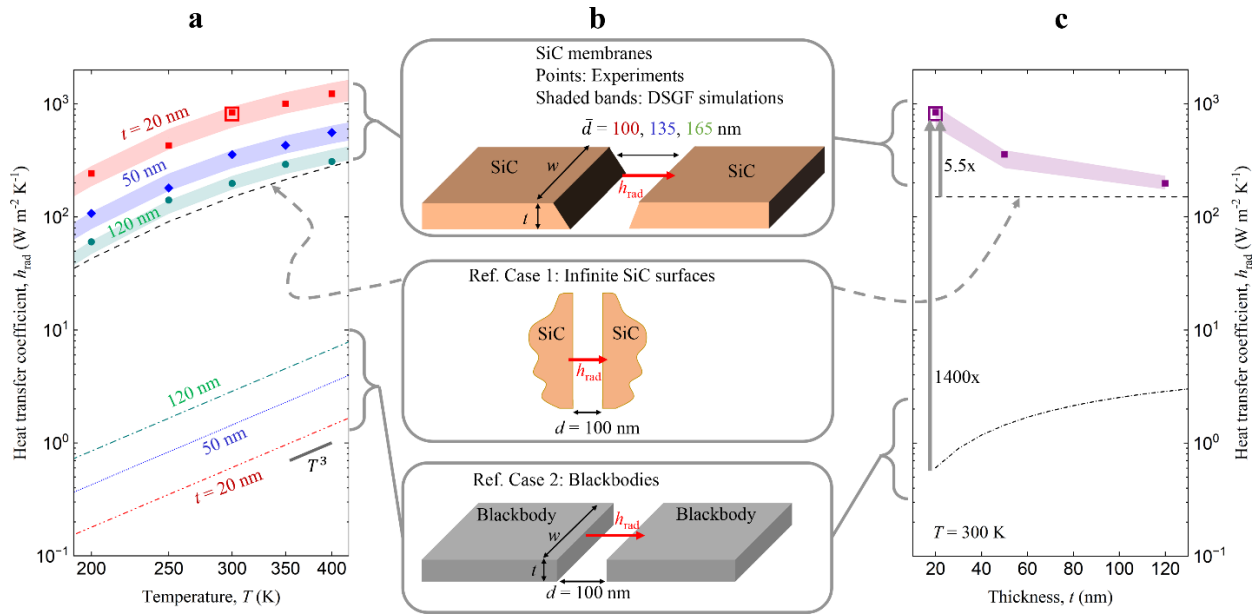

**Fig. 2 | Measured and modeled radiative heat transfer coefficients between coplanar SiC membranes. a,** Measured (solid-colored points) heat transfer coefficients for the three main

devices, denoted by their membrane thicknesses $t$ = 20, 50, and 120 nm, showing very good agreement with the corresponding DSGF simulations with no free parameters (color-coded shaded bands). The hollow red point is from another $t$ = 20 nm, $d$ = 100 nm device, measured only at 300 K. Each experimental point has an estimated $h_{rad}$ uncertainty of ±10% (omitted for visual clarity). The dashed lines are calculations for two reference cases: Ref. Case 1 is the near-field $h_{rad}$ between two infinite SiC surfaces separated by a gap $d$ = 100 nm, and Ref. Case 2 is $h_{rad}$ between parallel rectangular blackbodies evaluated using standard view factors corresponding to $d$ = 100 nm and the same $w$ and $t$ values as the three measured devices. **b,** Schematics of the device membrane geometries as well as the two reference cases. **c,** A slice through the data from **a** at 300 K, to better reveal how $h_{rad}$ increases as the membrane thickness $t$ decreases.

**Physical basis of the NFRHT enhancement in the dual nanoscale regime**

The large enhancement of NFRHT in the dual nanoscale regime fundamentally arises from the tunneling of evanescent electromagnetic corner and edge modes between the SiC membranes, a mechanism which is forbidden by symmetry[5-7,12-18] or negligible[8,9] in previous works. These modes are generated by the electromagnetic field couplings in the $x$-$z$ plane. There is no field coupling along the $y$-direction, because the membrane width $w$ is much larger than the thermal photon wavelengths (practically $w \to \infty$). As such, the emergence of corner and edge modes can be understood using a 2D structure as shown in Fig. 1a and Fig. 3, consistent with Berini's framework[10,11]. Specifically, in a solitary membrane the evanescent electromagnetic fields in the $x$-$z$ plane associated with surface phonon-polaritons (SPhPs) propagating along the $y$-direction couple between the membrane's perpendicular edges through the corners and between the adjacent corners to form four fundamental electromagnetic corner and edge modes[10,11]. Higher order resonances are also supported, but only the fundamental modes are important and therefore considered hereafter.

Figure 3a shows the spectral heat transfer coefficient at 300 K predicted using DSGF, along with a modal analysis of solitary 20, 50 and 120 nm thick SiC membranes in Fig. 3(b,c). The modal analysis is performed with COMSOL Multiphysics, and has been verified against the results of Ref. 11 for a silver membrane (see Supplementary Section 9). The modal analysis is performed on infinitely wide ($w \to \infty$), 2D membranes characterized by perfectly straight corners. Similarly, the spectral heat transfer coefficient is calculated for a uniform gap spacing of 100 nm (the calculations were repeated using nonuniform gap spacings, which was found to have almost no impact; see Supplementary Section 10).

**a**

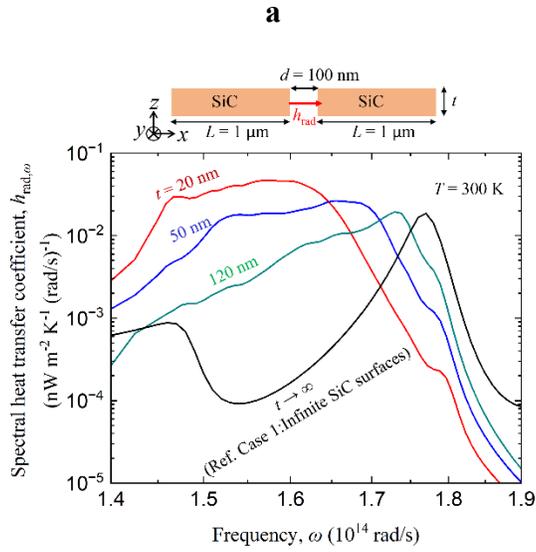

**b**

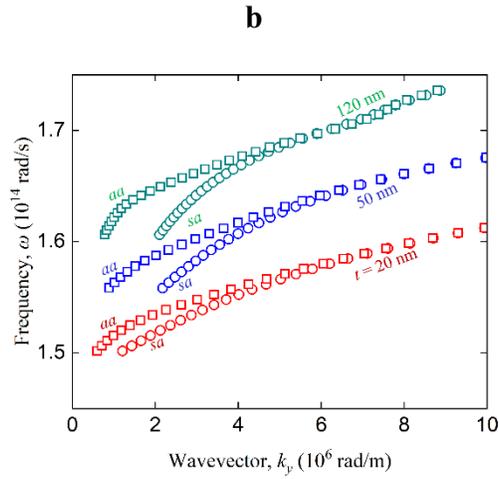

**c**

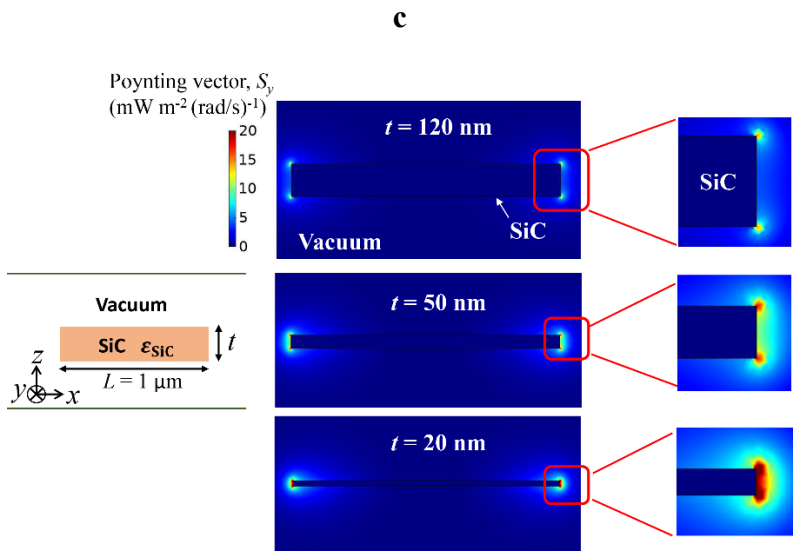

**Fig. 3 | Theoretical analysis of NFRHT enhancement mediated by electromagnetic corner and edge modes in the dual nanoscale regime between two coplanar SiC membranes.** DSGF simulations are performed on 3D membranes (finite $t$, $w$, and $L$), whereas COMSOL Multiphysics simulations are performed on 2D membranes (equivalent to $w \rightarrow \infty$). Both approaches here assume perfectly straight corners. **a,** DSGF calculated spectral heat transfer coefficient $h_{\mathrm{rad},\omega}$ at 300 K for membrane thicknesses of 20, 50, and 120 nm, as well for infinite parallel surfaces, all for a 100 nm gap. As the membrane thickness decreases, the resonance peak of $h_{\mathrm{rad},\omega}$ increases, redshifts, and broadens, which also increases the total NFRHT. **b,** Dispersion relations of the two low-frequency fundamental electromagnetic corner and edge modes, denoted *aa* and *sa* (see main text), in solitary 20, 50, and 120 nm thick membranes calculated with COMSOL Multiphysics. For each thickness, the *aa* and *sa* branches become degenerate with each other above $k_y \sim 5 \times 10^6$ rad/m. **c,** Poynting vector along the $y$-direction, $S_y$, calculated with COMSOL Multiphysics, showing representative electromagnetic corner and edge modes in solitary 20, 50 and 120 nm thick membranes. The Poynting vector of each membrane is evaluated at the frequency $\omega$ corresponding to $k_y = 5 \times 10^6$ rad/m, as determined using the *aa* mode dispersion relations of **b**.

For vacuum gap spacings in the deep subwavelength regime like in the present work ($d \ll \lambda_{\mathrm{th}}$), NFRHT between two infinite SiC surfaces is known to be dominated by SPhPs confined to the neighborhood of a single SiC-vacuum interface; these SPhPs can exist only in TM polarization[2,3]. The SPhP resonant frequency of a planar SiC-vacuum interface is estimated to be $1.77 \times 10^{14}$ rad/s from a SPhP dispersion relation (see Methods), which is in good agreement with the high-frequency peak in the spectral heat transfer coefficient calculated between two infinite surfaces (Fig. 3a). Compared to the case of two infinite surfaces, DSGF calculations of the spectral heat transfer coefficient in Fig. 3a for nanosize $t$ show that the resonance is redshifted and broadened as the membrane thickness decreases.

This trend is physically explained by analyzing the electromagnetic corner and edge mode dispersion relations (Fig. 3b), for the angular frequency, $\omega$, as a function of the real component of the wavevector along the $y$-axis, $k_y$. The four fundamental corner and edge modes are categorized in terms of the symmetry (*s*) and asymmetry (*a*) of the *z*-component of the electric field with respect to the *z* and *x* axes, respectively[11]. Here, we focus on the two low-frequency modes, designated *aa* and *sa*, corresponding to two branches in the dispersion relations, because only these modes are important for the resonances much below the longitudinal optical phonon frequency ($1.801 \times 10^{14}$ rad/s, see Methods); the *as* and *ss* modes are not important in this regime. Further, for every thickness considered, the calculations in Fig. 3b show that for $k_y$ above $\sim 5 \times 10^6$ rad/m, the *aa* and *sa* modes become essentially degenerate with each other and merge into a single branch in the dispersion relations. Also, for the 120 nm thick membrane, the dispersion relation reached an asymptote of nearly constant $\omega$ for $k_y > 9 \times 10^6$ rad/m, beyond which the COMSOL Multiphysics calculations became infeasible.

The thin-membrane geometries of the present work can sustain much richer electromagnetic modes than traditional infinite surfaces. For example, while the confined SPhPs at a single SiC-vacuum interface are purely TM-polarized, the corner and edge modes in membranes are not, since all six field components exist for all modes[11] (see Supplementary Section 11). In the electrostatic limit (i.e., $k_y$ much larger than the vacuum wavevector, $k_0$), for polar dielectrics like SiC, the largest contributing wavevector dominating NFRHT is estimated as $k_y \approx d^{-1}$,[40] which for a 100-nm vacuum gap spacing corresponds to $k_y \approx 1.0 \times 10^7$ rad/m. Now referring back to Fig. 3b for the corresponding $\omega$, we can expect the spectral $h_{\text{rad},\omega}$ to peak at around $1.61 \times 10^{14}$, $1.68 \times 10^{14}$ and $1.74 \times 10^{14}$ rad/s for the 20, 50 and 120 nm thick membranes, respectively. Finally, inspection of the actual calculated spectral $h_{\text{rad},\omega}$ curves in Fig 3a confirms that all three do indeed peak at frequencies close to the values deduced from Fig. 3b. This excellent agreement between the resonance predictions of Fig. 3b and the spectral heat transfer coefficient peaks in Fig. 3a, obtained using two entirely different calculations, explains the resonance redshift observed as the membrane thickness decreases. Also, the resonance broadening with decreasing thickness seen in Fig. 3a can be explained by the fact that losses in SiC increase within the Reststrahlen spectral band as the frequency is redshifted (see Supplementary Section 12 for details).

Turning to a representative real space visualization, Fig. 3c shows an example of the $y$-component of the Poynting vector, $S_y$, for three different membranes for the $aa$ mode at $k_y = 5 \times 10^6$ rad/m. For the 120 nm thick membrane, this Poynting energy is relatively small and is localized to the corners of the membrane. Then as the membrane thickness decreases the modes couple more strongly in the $z$-direction, causing $S_y$ to become larger around both the corners and vertical edges, which is what allows these electromagnetic corner and edge modes to dominate NFRHT for the 50 and 20 nm thick membranes.

To quantify this relative contribution of corner and edge modes to the NFRHT, above and beyond the traditional NFRHT between infinite planar surfaces, we introduce a NFRHT enhancement ratio: $R = \frac{h_{\mathrm{rad}}(d,t)}{h_{\mathrm{rad}}(d,t\to\infty)}$, where $h_{\mathrm{rad}}(d,t)$ is for a thin pair of coplanar membranes and $h_{\mathrm{rad}}(d,t\to\infty)$ is for infinite surfaces at the same gap spacing $d$. In both cases we assume $w \gg \lambda_{\mathrm{th}}$ so that practically $w \to \infty$.

Calculations for $t = 20$ nm and a uniform gap $d = 100$ nm at 300 K yield $R = 5.3$, indicating that $\frac{R-1}{R} = 81\%$ of the total NFRHT is mediated by the corner and edge modes. Similar calculations for 50 nm thick membranes with a 100-nm gap give $R = 3.4$, so that corner and edge modes still dominate with a relative contribution of 71%. Finally, calculations for $t = 120$ nm, $d = 100$ nm yield $R = 1.7$, showing that now more than half (59%) of the NFRHT is instead mediated by the SPhPs that already exist in infinite planes (here corresponding to the $y$-$z$ plane) with their evanescent fields decaying in the $x$-direction. Clearly, for ever thicker membranes ($t \to \infty$), the relative impact of corner and edge modes must become ever more negligible, eventually reverting to the NFRHT between two infinite plates ($R \to 1$).

In summary, we predicted and measured NFRHT in the dual nanoscale regime between two coplanar SiC membranes with thickness comparable to or smaller than their vacuum gap spacing of ~100 nm for temperatures ranging from 200 to 400 K. The measurements agree very well with theoretical predictions based on the DSGF method and no free parameters. The results demonstrate, for the first time, a novel mechanism mediated by electromagnetic corner and edge modes for enhancing NFRHT well beyond that between two infinite parallel surfaces. These newly-observed resonant modes can dominate the NFRHT with a relative contribution exceeding 80% for the thinnest membranes in this study. The measured radiative heat transfer coefficient at 300 K between 20 nm thick SiC membranes is 830 W/m$^2$K, which is 5.5 and 1400 times larger than that between two infinite SiC planes and two blackbodies, respectively. The emergence of electromagnetic corner and edge resonant modes in subwavelength membranes opens a new route for enhancing and spectrally controlling NFRHT. Such high heat transfer could enable future applications of NFRHT in noncontact localized radiative cooling, thermal management, and energy conversion devices.

**Methods**

**DSGF simulations of NFRHT.** NFRHT between two SiC membranes separated by a vacuum gap is modeled using the DSGF method[32], which is a numerically exact volume integral formulation based on fluctuational electrodynamics[20]. The DSGF method calculates the unknown system Green's function, $\overline{\overline{\mathbf{G}}}(\mathbf{r}, \mathbf{r}', \omega)$, relating the electric field with frequency $\omega$ observed at point $\mathbf{r}$ to a point source excitation at $\mathbf{r}'$. The unknown system Green's function is expressed in terms of the known free-space Green's function, $\overline{\overline{\mathbf{G}}}^0(\mathbf{r}, \mathbf{r}', \omega)$, as follows:

$$\overline{\overline{\mathbf{G}}}^0(\mathbf{r}, \mathbf{r}', \omega) = \overline{\overline{\mathbf{G}}}(\mathbf{r}, \mathbf{r}', \omega) - k_0^2 \int_{V_{\text{therm}}} \overline{\overline{\mathbf{G}}}^0(\mathbf{r}, \mathbf{r}'', \omega)[\varepsilon(\mathbf{r}'', \omega) - 1]\overline{\overline{\mathbf{G}}}(\mathbf{r}'', \mathbf{r}', \omega) \, d^3\mathbf{r}'' \qquad (1)$$

where $k_0$ is the vacuum wavevector magnitude, $\varepsilon$ is the dielectric function of SiC, and $V_{\text{therm}}$ is the combined total volume of the thermal objects (here, the two membranes). The free-space Green's function in vacuum is calculated as:

$$\overline{\overline{\mathbf{G}}}^0(\mathbf{r}, \mathbf{r}', \omega) = \frac{\exp(ik_0 r)}{4\pi r}\left[\left(1 - \frac{1}{(k_0 r)^2} + \frac{i}{k_0 r}\right)\overline{\overline{\mathbf{I}}} - \left(1 - \frac{3}{(k_0 r)^2} + \frac{3i}{k_0 r}\right)\hat{\mathbf{r}}\hat{\mathbf{r}}^\dagger\right] \qquad (2)$$

where $r = |\mathbf{r} - \mathbf{r}'|$, $\hat{\mathbf{r}} = \frac{\mathbf{r} - \mathbf{r}'}{|\mathbf{r} - \mathbf{r}'|}$, $\overline{\overline{\mathbf{I}}}$ is the unit dyadic, and $\dagger$ is the conjugate transpose operator. The volume of the thermal objects, $V_{\text{therm}}$, is discretized into $N$ cubic subvolumes along a cubic lattice. From Eq. (1), a system of linear equations of the form $\overline{\overline{\mathbf{A}}}\overline{\overline{\mathbf{G}}} = \overline{\overline{\mathbf{G}}}^0$ is derived,[32] where $\overline{\overline{\mathbf{A}}}$ is the $3N \times 3N$ interaction matrix accounting for $N$ subvolumes and three Cartesian components, $\overline{\overline{\mathbf{G}}}^0$ is the $3N \times 3N$ matrix containing the discretized free-space Green's function, and $\overline{\overline{\mathbf{G}}}$ is the $3N \times 3N$ matrix containing the unknown system Green's function. Once the system of linear equations is solved, the total radiative heat transfer coefficient, $h_{\text{rad}}$, is calculated as:

$$h_{\text{rad}}(T) = \frac{1}{2\pi A_c} \int_0^\infty \left.\frac{\partial \Theta(\omega, T')}{\partial T}\right|_{T'=T} \tau(\omega) d\omega \qquad (3)$$

where $\Theta(\omega, T)$ is the mean energy of an electromagnetic state and $A_c = wt$ is the membrane cross-section. The spectral transmission coefficient between the two membranes of volumes $V_h$ and $V_c$, denoted $\tau(\omega)$, is calculated from the system Green's function:

$$\tau(\omega) = \sum_{i \in V_h} \sum_{j \in V_c} 4k_0^4 \Delta V_i \Delta V_j \, \text{Im}[\varepsilon(\mathbf{r}_i, \omega)] \text{Im}[\varepsilon(\mathbf{r}_j, \omega)] \text{Tr}[\overline{\overline{\mathbf{G}}}(\mathbf{r}_i, \mathbf{r}_j, \omega)\overline{\overline{\mathbf{G}}}^\dagger(\mathbf{r}_i, \mathbf{r}_j, \omega)] \qquad (4)$$

where $\Delta V_i$ and $\Delta V_j$ are respectively the volumes of the $i^{\text{th}}$ and $j^{\text{th}}$ subvolumes used in the discretization. The spectral heat transfer coefficient shown in Fig. 3a is obtained by evaluating Eq. (3) without the integration over $\omega$. The contribution from propagating waves to NFRHT is calculated by considering only the far-field (FF) term in the free-space Green's function (see Supplementary Section 7)[41]:

$$\overline{\overline{\mathbf{G}}}^{0,\text{FF}}(\mathbf{r}, \mathbf{r}', \omega) = \frac{\exp(ik_0 r)}{4\pi r}[1 - \hat{\mathbf{r}}\hat{\mathbf{r}}^\dagger] \qquad (5)$$

**NFRHT between two infinite surfaces (Reference Case 1).** The NFRHT between two SiC infinite surfaces (Reference Case 1 of Fig. 2 and Fig. 3a) is calculated using a closed-form expression derived from fluctuational electrodynamics. The propagating and evanescent components of the total radiative heat transfer coefficient are respectively calculated as[2]:

$$h_{\text{rad,inf}}^{\text{prop}}(T) = \frac{1}{4\pi^2} \int_0^\infty d\omega \left.\frac{\partial \Theta(\omega, T')}{\partial T}\right|_{T'=T} \int_0^{k_0} dk_\rho k_\rho \sum_{\gamma = TE, TM} \frac{(1-|r_{0h}^\gamma|^2)(1-|r_{0c}^\gamma|^2)}{|1 - r_{0h}^\gamma r_{0c}^\gamma e^{2i\text{Re}(k_{x0})d}|^2} \qquad (6)$$

$$h_{\text{rad,inf}}^{\text{evan}}(T) = \frac{1}{\pi^2} \int_0^\infty d\omega \left.\frac{\partial \Theta(\omega, T')}{\partial T}\right|_{T'=T} \int_{k_0}^\infty dk_\rho k_\rho e^{-2\text{Im}(k_{x0})d} \sum_{\gamma = TE, TM} \frac{\text{Im}(r_{0h}^\gamma)\text{Im}(r_{0c}^\gamma)}{|1 - r_{0h}^\gamma r_{0c}^\gamma e^{-2\text{Im}(k_{x0})d}|^2} \qquad (7)$$

where the subscripts h and c respectively refer to the hot and cold surfaces, whereas 0 denotes vacuum. In Eqs. (6) and (7), $k_\rho$ is the wavevector parallel to the $y$-$z$ plane, $k_{x0}$ is the $x$-component of the vacuum wavevector normal to the surfaces, and $r_{0h}^\gamma$ and $r_{0c}^\gamma$ are respectively the Fresnel reflection coefficients at the vacuum-hot and vacuum-cold interfaces in polarization state $\gamma$. The total heat transfer coefficient reported for Reference Case 1 in Fig. 2 is the sum of Eqs. (6) and (7). Similarly, the spectral heat transfer coefficient for Reference Case 1 in Fig. 3a is the sum of Eqs. (6) and (7) without the integration over $\omega$.

**Dielectric function of polycrystalline SiC.** The temperature-independent dielectric function of polycrystalline SiC is described by a Lorentz model with the parameters taken from Ref. 8:

$$\varepsilon(\omega) = \varepsilon_\infty \frac{\omega^2 - \omega_{LO}^2 + i\Gamma\omega}{\omega^2 - \omega_{TO}^2 + i\Gamma\omega} \tag{8}$$

where $\varepsilon_\infty = 8$, $\omega_{LO} = 1.801 \times 10^{14}$ rad/s, $\omega_{TO} = 1.486 \times 10^{14}$ rad/s, and $\Gamma = 3.767 \times 10^{12}$ rad/s are respectively the high-frequency dielectric constant, the longitudinal optical phonon frequency, the transverse optical phonon frequency, and the damping constant.

**Radiative heat transfer between two black membranes (Reference Case 2).** The total radiative heat transfer coefficient between two membrane edges in the blackbody limit is calculated via the Stefan-Boltzmann law[39]:

$$h_{rad}^{BB}(T) = 4\sigma_{SB}T^3 F_{12} \tag{9}$$

where $\sigma_{SB}$ is the Stefan-Boltzmann constant and $F_{12}$ is the view factor between the membrane edge face areas (treated as parallel rectangles of dimensions $w \times t$) calculated at a gap spacing $d$ of 100 nm[39], which is the fraction of radiation energy leaving the first rectangle intercepted by the second. The view factor approaches unity when the membrane width $w$ and thickness $t$ are both very large compared to $d$ ("infinite surfaces"). Note that this $h_{BB}$ expression also explains the $T^3$ power law seen for Reference Case 2 in Fig. 2a, which fundamentally arises from the specific heat of photons in a 3D gray medium.

**SPhP resonance of a SiC-vacuum interface.** NFRHT between two infinite SiC surfaces in the deep subwavelength regime is quasi-monochromatic at the SPhP resonant frequency of a single SiC-vacuum interface. This resonant frequency, derived from a SPhP dispersion relation for a SiC-vacuum interface, can be approximated as[40]:

$$\omega_{res,inf} \approx \sqrt{\frac{\varepsilon_\infty \omega_{LO}^2 + \omega_{TO}^2}{\varepsilon_\infty + 1}} \tag{10}$$

The calculated $\omega_{res,inf}$ of $1.77{\times}10^{14}$ rad/s is in excellent agreement with the maximum of the spectral heat transfer coefficient for two infinite SiC surfaces (see Fig. 3a).

**Uncertainty analysis.** *Theoretical predictions.* The shaded bands of theoretical predictions in Fig. 2 account for the uncertainties of the gap spacing from the SEM images as well as the tilting of the membranes from the LCM measurements. The estimated uncertainty of the gap spacing is ±10 nm. The estimated tilt angles are 0.8°, 4.8°, and 3.4° for the 120, 50, and 20 nm thick devices, respectively (see Supplementary Section 2). The total heat transfer calculated by DSGF decreases when the membranes are tilted and when the separation gap increases. Therefore, as an example, the upper bound of the red-shaded band in Fig. 2a for the 20 nm thick device is obtained from a separate DSGF calculation assuming no tilt and by using the lower-bound estimated gap spacing of 90 nm (found by subtracting the estimated uncertainty of 10 nm from the nominal gap spacing of 100 nm). Similarly, the lower bound of the red-shaded band comes from another DSGF calculation using the full tilting of 3.4° and by assuming an upper-bound gap of 110 nm. The same procedure is used to calculate the theoretical curves for the 50 and 120 nm thick devices, each based on their respective estimated gap spacings, tilts, and uncertainties thereof. Note that the effects of tilting on NFRHT between the SiC membranes are small; for example, $h_{rad}$ decreases by only 11% for the maximum tilt angle of 4.8° in the 50 nm thick device.

In addition, the 120, 50, and 20 nm thick SiC membranes may be misaligned by respectively 20, 10, and 10 nm along the $z$-direction according to the LCM line scans (see Supplementary Section 2). DSGF simulations show that such misalignments have a negligible impact on the total radiative heat transfer, such that they are not included in the uncertainty analysis. For example, $h_{rad}$ decreases by less than 1% when the 20 nm thick membranes are misaligned by 10 nm in $z$.

*Experimental data.* The estimated uncertainty of the measured heat transfer coefficients in Fig. 2 is ±10%, which comes from the uncertainties of the membrane dimensions and in the resistance thermometry of the PRTs. The widths and thicknesses, respectively, of the membranes are determined using a Zeiss SEM and a spectroscopic ellipsometer (Angstrom Sun Technologies model SE200BM), both with uncertainties estimated as 2% or better (see also Ref. 33). The uncertainties in the resistance thermometry are typically 5% or less of the relevant $\Delta T$ (see Supplementary Sections 4 and 5).

**Data availability.** The data that support the findings of this study are available from the corresponding authors on reasonable request.

**Acknowledgments.** L.T. and C.D. gratefully acknowledge financial support from the Marjorie Jackson Endowed Fellowship Fund, the Howard Penn Brown Chair, and the Army Research Laboratory as part of the Collaborative for Hierarchical Agile and Responsive Materials (CHARM) under Cooperative Agreement No. W911NF-19-2-0119. L.M.C. and M.F. acknowledge the financial support from the National Science Foundation (grant number CBET-1952210) and from the Natural Sciences and Engineering Research Council of Canada (funding reference number RGPIN-2023-03513). This work was partially performed at the UC Berkeley Marvell Nanolab. The authors appreciate the support of the staff and facilities that made this work possible. This research used the Savio computational cluster resource provided by the Berkeley Research Computing program at the University of California, Berkeley (supported by the UC Berkeley Chancellor, Vice Chancellor for Research, and Chief Information Officer).

**Author contributions.** This work was initialized and conceived by L.T. Design, fabrication and testing of the devices were performed by L.T. under the supervision of C.D. Numerical simulations were performed by L.T. with help of L.M.C. under the supervision of M.F. All authors contributed to the data analysis. The manuscript was written by L.T., M.F. and C.D.

**Competing interests.** The authors declare no competing interests.

**Supplementary Information**

**Large enhancement of near-field radiative heat transfer in the dual nanoscale regime enabled by electromagnetic corner and edge modes**

Lei Tang[1], Lívia Corrêa[2], Mathieu Francoeur[3,*], and Chris Dames[1,*]

[1]Department of Mechanical Engineering, University of California at Berkeley, Berkeley, CA, USA
[2]Department of Mechanical Engineering, University of Utah, Salt Lake City, UT, USA
[3]Department of Mechanical Engineering, McGill University, Montréal, QC, Canada
[*]Corresponding authors: mathieu.francoeur@mcgill.ca, cdames@berkeley.edu

<u>**Contents**</u>

**1. Device fabrication**

**2. Characterization of membrane coplanarity**

**3. Asymmetric "knife edge" imaging artifacts in laser confocal microscope (LCM)**

**4. Heat transfer measurements**

**5. Parasitic heat transfer measurements**

**6. Convergence of DSGF simulations**

**7. Contribution from propagating waves to NFRHT between two SiC membranes**

**8. Additional thickness-dependent comparisons between experiments and DSGF simulations**

**9. Verification of modal analysis**

**10. Comparison of spectral radiative heat transfer coefficients calculated based on uniform and nonuniform gap spacings**

**11. Spatial distribution of electric and magnetic fields of the asymmetric-asymmetric (*aa*) mode**

**12. Dielectric function of polycrystalline SiC**

# 1. Device fabrication

The devices used for measuring near-field radiative heat transfer (NFRHT) between two SiC membranes are similar to those described in our previous work[1] and are manufactured using standard microfabrication technique as shown in Supplementary Fig. 1 and detailed here.

A SiC layer of the desired thickness is first grown on a Si substrate using low-pressure chemical vapor deposition (LPCVD), and is then immediately moved to a nitrogen ambient furnace for thermal annealing for 2 hours at 1100°C (Step 1). In Step 2, the SiC layer is patterned using chlorine-based reactive ion etching (RIE), with a home-developed recipe (the flow rates of $Cl_2$, HBr, and $CF_4$ are respectively 125, 75, and 100 sccm, and the RF power is 300 W). For the 20 and 50 nm thick devices, the central portion of the SiC layer that will become the membranes in the finished device is initially patterned to be 30 μm wide ($y$-direction) and 4 μm long ($x$-direction); for the 120 nm thick device, this central section is 20 μm long. The side portions of the SiC layer that will become the wings in the finished device have the same width, but half of the length, of the middle portion. In Step 3, a low-stress $SiN_x$ film is grown via LPCVD and pattered using fluorine-based RIE. The thickness of $SiN_x$ is 500 nm for the 120 nm thick device and is 200 nm for the 50 and 20 nm thick devices. In order to minimize the thermal contact resistances, the $SiN_x$ is patterned to slightly overlap with all SiC contact edges by 1 μm. Therefore, at this point, the exposed central portion of the SiC film is 2 or 18 μm long (i.e., 2 μm shorter than the 4 or 20 μm underlying portion patterned in Step 2). Thus after cutting the central SiC in half to define the gap $d$, the final released $L$ as depicted in Fig. 1b of the main text will be around 1 μm for the 20 and 50 nm thick devices and 9 μm for the 120 nm thick device. The same $SiN_x$ overlap and SiC length reduction of 1 μm is also applied for the two side portions (wings) of the SiC film. Then in Step 4 the serpentine Pt resistance thermometers (PRTs) and their associated electrical connections are manufactured by patterning a 50 nm thick Pt film along with a 5 nm Cr adhesion layer using a lift-off process.

Next, the most challenging fabrication procedure is to create the nanosized gap spacing in the central portion of the SiC film to obtain two separate membranes for NFRHT measurements. This is accomplished via the following steps. A thin layer of $SiO_2$ hard mask is first grown (Step 5) using plasma-enhanced chemical vapor deposition (PECVD). The thicknesses of $SiO_2$ are 200, 100, and 50 nm for the 120, 50, and 20 nm thick devices, respectively. Then, a thin layer of Zep 520 e-beam resist (200 - 300 nm thick) is spin-coated and patterned by a Crestec CABL series ultra-high resolution electron beam nanolithography system. The pattern is transferred to the $SiO_2$ by an STS advanced planar source oxide etch system with a high coil power for better sidewall profile (Step 6). Then in Step 7 the e-beam resist is removed and the $SiO_2$ pattern is transferred to the SiC using chlorine-based RIE with the same recipe as described in Step 2. Finally, after removing the extra $SiO_2$, the device is released in a KOH solution at 80°C for at least 2 hours. Additional details regarding device fabrication can be found in Ref. 1.

Note that the fourth device with $t$ = 20 nm and $L$ = 2 μm for the additional verification mentioned in the main text is fabricated using the same process as the device with $t$ = 20 nm and $L$ = 1 μm. The only difference is that the central SiC section is initially patterned to be 6 μm long rather than 4 μm long in Step 2; the resulting final released $L$ is thus 2 μm.

Regarding the KOH release in Step 8, we emphasize that the etching of SiC in the KOH is negligible since SiC is known to be highly resistant to KOH, and indeed SiC has been used as a protective layer or hard mask for the KOH etching of Si[2,3]. For verification, we also prepared a test sample with a 120 nm thick SiC film deposited onto a Si substrate and submerged it into the KOH solution at the same temperature for the same period of time used for releasing the main devices. We then measured and compared the thickness of this test SiC film before and after the KOH immersion. The thicknesses of the SiC film before and after the KOH was essentially the same, with a difference less than 1%, i.e., less than 1 nm, thereby confirming that the KOH release protocol of Step 8 has a negligible impact on etching the SiC.

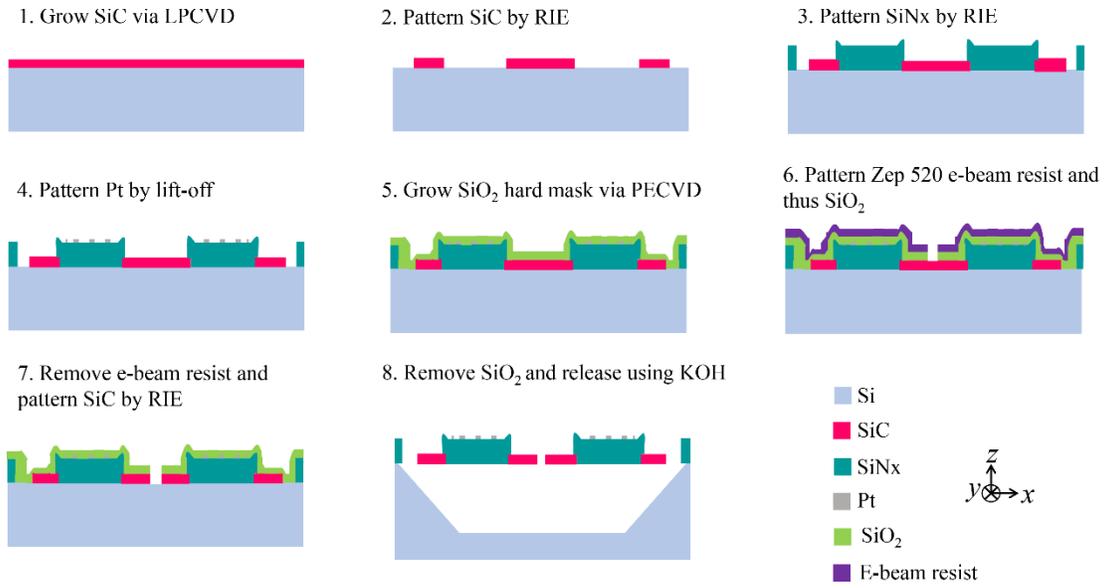

**Supplementary Fig. 1 | Device fabrication.** Main steps for fabricating the suspended devices used for measuring NFRHT between two SiC membranes.

## 2. Characterization of membrane coplanarity

As mentioned in the main text, the membrane coplanarity of each finished device after the release etch is characterized with an Olympus LEXT OLS4000 3D laser confocal microscope (LCM). The LCM data for the 20, 50, and 120 nm thick devices (three main devices) are shown in Supplementary Fig. 2. Note that for the 120 nm thick device the LCM line scans are unreliable for $x$ values within around ±1 µm of the gap spacing due to "knife edge" optical artifacts and were thus discarded (see Supplementary Section 3 for details).

The membranes are near-coplanar based on the LCM scanning data at three different line scan locations (marked as 1, 2 and 3 in the inset of Supplementary Fig. 2). The membrane tilt angles $\theta$ are respectively 3.4°, 4.8°, and 0.8° for the 20, 50, and 120 nm thick devices. Based on discrete system Green's function (DSGF) simulations of NFRHT, we estimate that a membrane tilt of 4.8° decreases the total radiative heat transfer coefficient by 11% compared to that for the perfectly coplanar membranes. Such small reduction in NFRHT from the membrane tilt is mainly due to the increased gap size $d$, which can be estimated by $2L(1 - \cos\theta)$ (e.g., $d$ increases by 7 nm for the 50 nm thick device since $L = 1$ µm and $\theta = 4.8°$). We incorporated the potential impact of imperfect coplanarity through the breadth of the colored bands in the theoretical $h_{rad}$ presented in Fig. 2 (see Methods for details). Based on the data variations of three LCM line scans shown in Supplementary Fig. 2, we also estimated that the membrane misalignments along $z$ are approximately 10, 10, and 20 nm for the 20, 50, and 120 nm thick devices, respectively. Such misalignments have negligible impacts on the total radiative heat transfer according to additional DSGF simulations (see Methods).

The dimensions of four devices used for NFRHT measurements along with their characterization parameters of the gap spacing and membrane coplanarity are summarized in Supplementary Table 1.

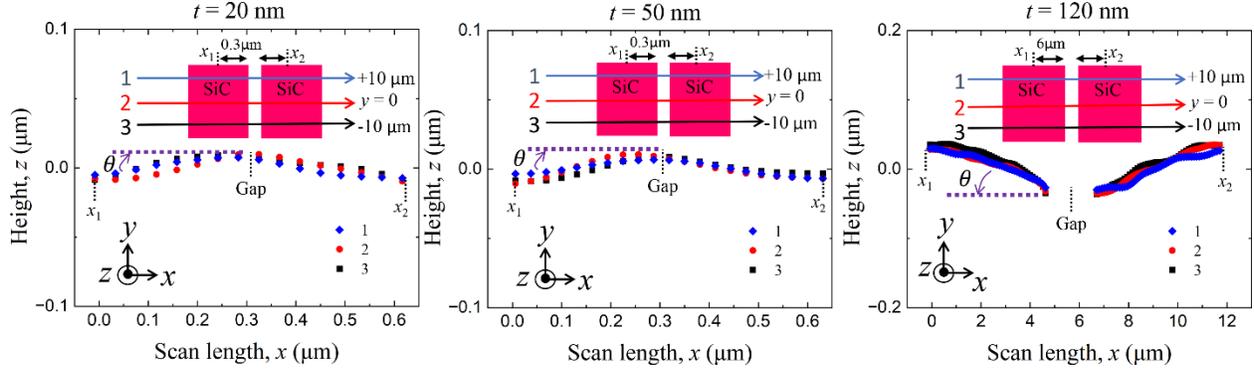

**Supplementary Fig. 2 | Characterization of the membrane coplanarity.** Laser confocal microscopy (LCM) line scans of flatness and co-planarity of the suspended SiC membranes of the three main devices after final release. The inset shows the locations of the three line-scan profiles. The tilt angles of the 20, 50, and 120 nm thick membranes respectively are $\theta$ = 3.4°, 4.8°, 0.8°. The 20, 50, and 120 nm thick membranes are misaligned by 10, 10, and 20 nm along $z$, respectively, according to the data variations of three LCM line scans for each device.

**Supplementary Table 1 | Details of the four devices used for NFRHT measurements.** The dimensions (thickness, $t$, and length, $L$), gap spacings ($d$ range, and average gap $\bar{d}$), tilt angles ($\theta$), and $z$-misalignments for the four pairs of SiC membranes are provided. All membranes have the same width of $w$ = 30 μm. The main devices are #1-3, with device #4 measured at 300 K.

| Device # | Thickness $t$ (nm) | Length $L$ (μm) | Gap $d$ and $\bar{d}$ (nm) | Tilt angle $\theta$ (degrees) | $z$-misalignment (nm) |
|---|---|---|---|---|---|
| 1 | 20 | 1 | $d \approx \bar{d} \approx 100$ | 3.4 | 10 |
| 2 | 50 | 1 | $d$ = 110 - 160 ($\bar{d} \approx 135$) | 4.8 | 10 |
| 3 | 120 | 9 | $d$ = 85 - 245 ($\bar{d} \approx 165$) | 0.8 | 20 |
| 4 | 20 | 2 | $d \approx \bar{d} \approx 100$ | 3.9 | 10 |

**3. Asymmetric "knife edge" imaging artifacts in laser confocal microscope (LCM)**

As mentioned in Supplementary Section 2, when performing the LCM line scan for the 120 nm thick device, we found that the data became unreliable for $x$ values within approximately $\pm 1$ μm of the gap and were thus excluded from the rightmost panel of Supplementary Fig. 2. This issue was observed only for the 120 nm thick sample, which we hypothesize arises because it was scanned with a different objective lens than the 20 and 50 nm thick samples; because the 120 nm sample has a much larger $L$ of suspended SiC (9 μm vs. 1 μm), an LCM objective with lower magnification and thus larger spot size was used in order to scan the much longer distance along the $x$-axis.

We attribute the unreliable LCM $z$ measurements of the 120 nm sample in the vicinity of the gap to optical "knife edge" artifacts near the two adjacent membrane edges: since the gap itself is essentially non-reflecting, when the LCM laser beam is moved close to the gap, the beam may be partially "cut" by the free SiC edges which results in inaccurate LCM measurements. We investigated this hypothesis by performing additional line scans around the free edge of the SiC wing located on the heating island side of the 120 nm thick device, i.e., the leftmost pink rectangle depicted in Fig. 1b of the main text; see also the inset of Supplementary Fig. 3. That SiC wing should have a similar knife edge and tilting angle as the SiC membrane adjacent to the gap, owing to the symmetry of the device fabrication, but as a single step it is easier to analyze by avoiding the complexity associated with the nanosize gap itself and the second adjacent suspended membrane. As shown in Supplementary Fig. 3 this additional LCM line scanning was performed two times at the same location on the same SiC wing, simply rotated at two different angles on the LCM stage (0° and 180° as shown in the inset). The LCM results for the free edge of this SiC wing up to ~1 μm along the $x$-axis obtained from the two different rotation angles are completely different from each other (circled data in the figure), which we consider non-physical since it is the same actual region of the SiC wing. Beyond that unreliable first ~1 μm of the edge, the red and blue data in Supplementary Fig. 3 are in very good agreement with each other, as expected, and yield a tilt angle of ~1° for the SiC wing (which, also as expected, is similar to the tilt angle of the central released SiC for that sample as listed in Supplementary Table 1). Therefore, the LCM measurements of this sample near the free SiC edges are clearly not accurate.

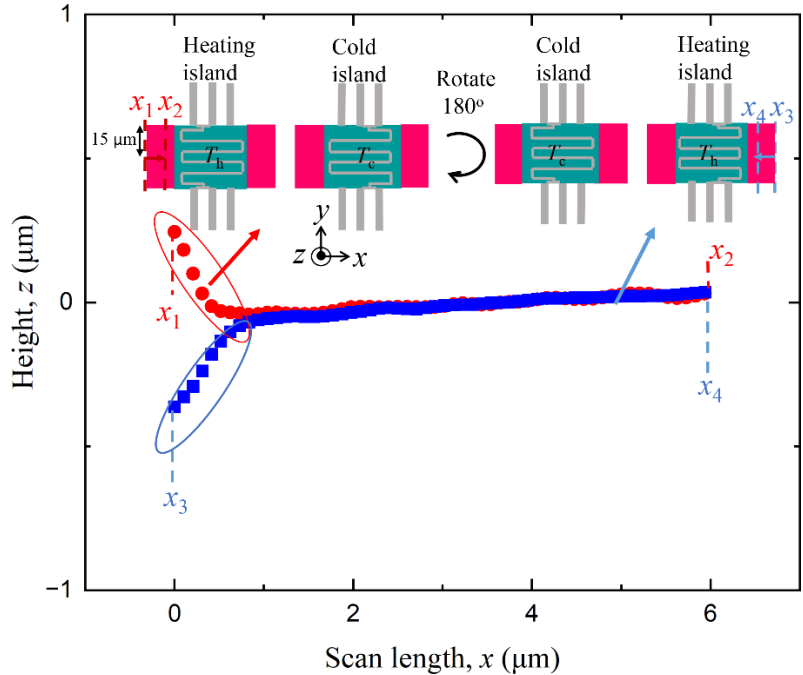

**Supplementary Fig. 3 | Knife edge effects from LCM when using the lower magnification lens.** LCM measurements for the SiC wing located on the heating island side of the 120 nm thick device for two different orientations. The red points are for a scan from $x_1$ to $x_2$, with the free end of the SiC wing at the left of the LCM stage. The blue points, from $x_3$ to $x_4$, is of the same SiC wing after rotating the sample 180° so that the free end of the wing is at the right of the LCM stage. While $x_1$ and $x_3$ refer to the same physical location on the wing, the corresponding red and blue LCM data clearly diverge from each other for the first ~1 µm along the *x*-axis and are not accurate there (labeled by two ellipses), which we attribute to knife edge effects.

## 4. Heat transfer measurements

For the heat transfer measurements, the microfabricated test device is thermally well anchored to the cryostat cold finger using thermal grease and spring screw pressure. AC lock-in methods are used to ensure high accuracy and sensitivity of the electrothermal measurements. An AC sinusoidal current, $I_f$, with typical amplitude up to 8 μA, at a frequency $f$ of 0.98 Hz is applied to the heating island. This leads to temperature rises in the heating and cold islands both at $2f$. The temperature rises of both islands need to be measured accurately in order to quantify the heat transfer between them. Note that the heat transfer thus measured includes NFRHT between the two SiC membranes as well as potential parasitic heat transfers which are discussed next in Supplementary Section 5. Following Refs. 4 and 5, the temperature rises at $2f$ of the heating island, $\Delta T_{h,2f}$, and the cold island, $\Delta T_{c,2f}$, are obtained by measuring the voltage fluctuations across the heating and cold islands' PRTs at $3f$ ($V_{h,3f}$) and $2f$ ($V_{c,2f}$) respectively (note the two different harmonics), and are calculated from:

$$\Delta T_{h,2f} = \frac{2V_{h,3f}}{I_f R_h \alpha_h} \tag{S1}$$

$$\Delta T_{c,2f} = \frac{V_{c,2f}}{I_{dc} R_c \alpha_c} \tag{S2}$$

where $I_{dc} = 10$ μA is the DC current supplied to the cold island, resulting in a maximum of only 3 K DC temperature rise. Also $R_{h,c}$ and $\alpha_{h,c}$ are respectively the electrical resistance and the temperature coefficient of resistance (TCR) of the heating island's (h) or the cold island's (c) PRT, as determined in a separate calibration step.

For the PRT calibration, the temperature dependent $R_h$ and $R_c$ are obtained by supplying a small AC current of amplitude $I_f = 1$ μA and frequencies of 787.7 Hz and 656.6 Hz, respectively, to each PRT. This results in negligible self-heating, and the voltage signal across each PRT at frequency $1f$ ($V_{1f}$) is then measured using a four-point scheme. Then, $R_h$ and $R_c$ are obtained by taking the ratio of $V_{1f}$ to $I_f$. The difference between $R_h$ and $R_c$ of each device is very small (typically less than 0.3%). This is expected since the fabrication process was completed in the same batch for the PRTs of each device (i.e., they were fabricated simultaneously on a single Si wafer). In addition, the TCRs of the PRTs are calibrated for every device and island, and a similar excellent agreement between the TCR of the heating island's PRT and the TCR of the cold island's PRT for each device was observed, as expected. The obtained TCRs are comparable to values from the literature; details can be found in our previous work[1].

In order to validate the obtained experimental results, we also carefully verified the thermal cutoff frequency and repeatability of our measurements as detailed next.

## 4.1. Thermal cutoff frequency

The frequency response of the devices needs to be characterized in order to determine an appropriate thermal cutoff frequency: if the heating frequency is too high, there is not enough time for the heat to fully spread among the device and reach a quasi-static temperature profile. To characterize this cutoff, we performed an auxiliary study by varying the frequency $f$ of the AC sinusoidal current (here of amplitude $I_f = 5$ μA) supplied to the heating island. The resulting temperature rise of the heating island at each frequency is obtained and recorded by measuring the voltage fluctuation at $3f$ with Eq. (S1). Supplementary Fig. 4a shows the measured temperature rise of the heating island as a function of heating current frequency, $f$, for two different devices at 300 K. Note that the temperature rise is normalized by the maximum $\Delta T_{h,2f}$ (~0.5 K) obtained from the frequency sweep. Also, during this frequency sweep, we monitored and recorded the temperature response of the cold island for a few heating frequencies. The similar normalized temperature rise of the cold island is shown in Supplementary Fig. 4b. Clearly, thermal attenuation at both hot and cold islands is negligible for heating current frequencies $f$ lower than 1 Hz. Therefore, a frequency of 0.98 Hz was used in the heat transfer measurements because it mitigates the effects of thermal attenuation.

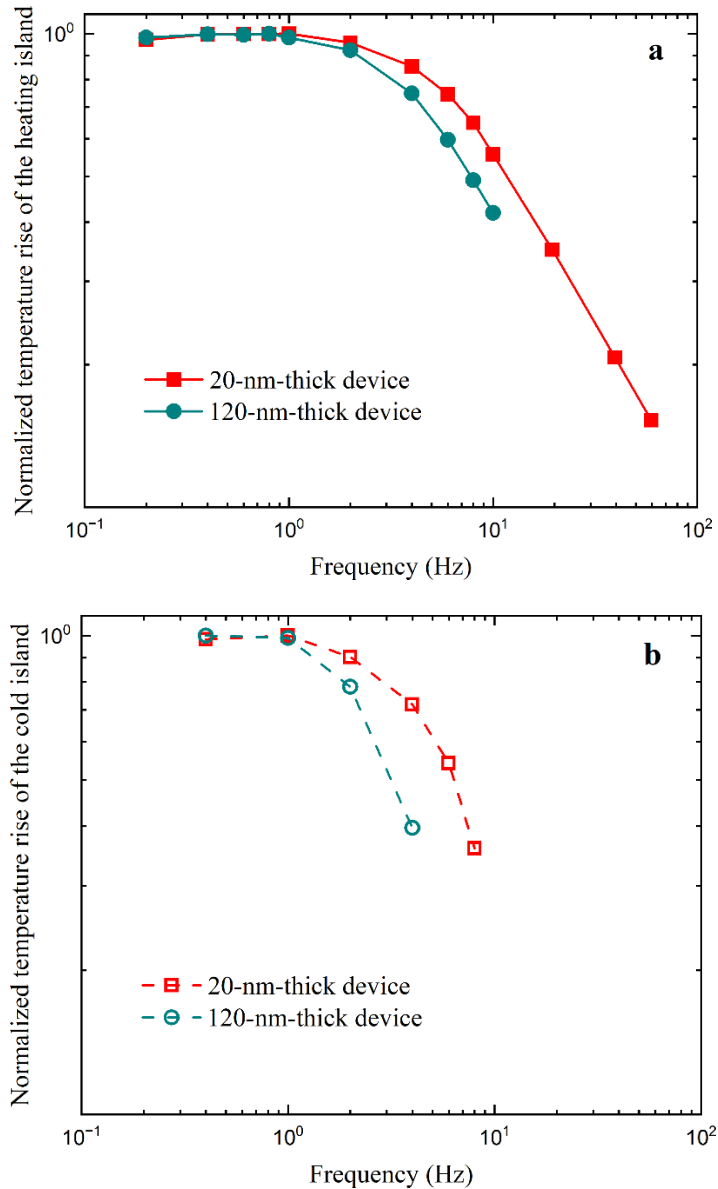

**Supplementary Fig. 4 | Frequency response of the 20 and 120 nm thick devices at 300 K.** The temperature rise, normalized by its maximum value, is plotted as a function of the heating current frequency, *f*. Thermal attenuation is negligible for frequencies lower than 1 Hz.

*4.2. Repeatability of the measurements*

To ensure the accuracy of the results, after calibrating the PRT *R*(*T*) curves, the NFRHT measurements are repeated by ramping the stage temperature in steps of 50 K from 200 K to 400 K, back down to 200 K, and finally back up to 400 K (thus there are three visits to each stage *T* from 250 K to 350 K, and two visits each at 200 K and 400 K). For every stage temperature and visit, after stabilizing, a complete NFRHT measurement set is taken as follows. The heating current $I_f$ is ramped in 0.5 µA steps in three phases: first from 1 µA to 8 µA, back down to 1 µA, and finally back up to 8 µA. A representative example of such a triply-redundant current sweep dataset is shown in Supplementary Fig. 5, demonstrating excellent repeatability in both the measured Joule heat $Q(I_f)$ and temperature rise $\Delta T_{tot,2f}(I_f)$. Note that each experimental $h_{rad}$ point shown in Fig. 2 of the main text corresponds to the average value of the data obtained from such repetitions of both stage $T$ and current sweep, with a typical variability of less than 5%.

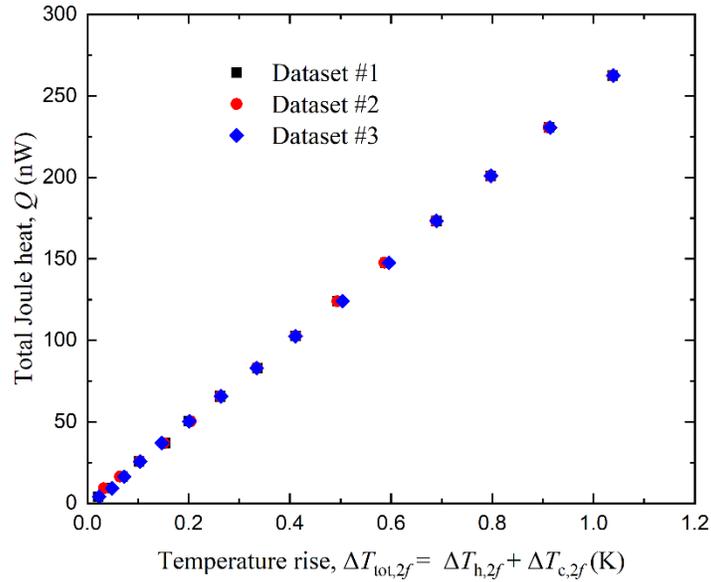

**Supplementary Fig. 5 | Repeatability of the measurements.** Representative example of total Joule heat versus temperature rise, here for the 20 nm thick device at a cryostat stage temperature of 300 K. Datasets (1, 2, 3) correspond to (increasing, decreasing, increasing) current sweeps.

## 5. Parasitic heat transfer measurements

In addition to the two SiC membranes between which the NFRHT occurs, each device includes other structures like the SiNx islands with PRTs and the SiNx supporting beams with Pt lines. Such structures introduce the potential for additional parasitic contributions to the heat transfer between the two islands, such as far-field radiation contributions between the islands either directly through their small but finite view factors or reflected off the surrounding substrate, or conduction pathways through the supporting beams and substrate frame. For every device, we quantify this parasitic heat transfer by performing measurements on a corresponding reference device. Every reference device has the same geometries and is prepared in the same microfabrication run as its corresponding device, the only difference being that the central SiC membranes were etched and removed from the reference device. The reference devices are fabricated using the process described in Supplementary Section 1, except that the patterns of the e-beam resist and thus $SiO_2$ in Step 6 shown in Supplementary Fig. 1 are different: the central portion of the resist and $SiO_2$ are removed in order to etch away the central SiC membranes.

An example of this background subtraction is given in Supplementary Fig. 6, for the 120 nm thick device. First, the total thermal conductance, $G_1$, is measured on a complete device that includes the central SiC membranes (Configuration 1, with geometry depicted in the figure inset). Then, the background thermal conductance, $G_2$, is obtained from similar measurements on the reference device from which the central SiC membranes had been etched and removed (Configuration 2 in the inset). Here, $G_2$ accounts for all heat transfer contributions that are not due to NFRHT between the central SiC membranes. Finally, the NFRHT conductance is calculated as $G_{\text{NFRHT}} = G_1 - G_2$.

For a quick graphical interpretation of Supplementary Fig. 6, it is helpful to also recognize that for our systems $\Delta T_{\text{h},2f} \gg T_{\text{c},2f}$ for both Configuration 1 and Configuration 2, for all devices and temperatures. With this simplification it can be shown that $\frac{G_2}{G_1} \approx \frac{\left[\frac{\partial \Delta T_{\text{c},2f}}{\partial \Delta T_{\text{h},2f}}\right]_2}{\left[\frac{\partial \Delta T_{\text{c},2f}}{\partial \Delta T_{\text{h},2f}}\right]_1}$, where the relevant slopes obtained from Supplementary Fig. 6 are $\left[\frac{\partial \Delta T_{\text{c},2f}}{\partial \Delta T_{\text{h},2f}}\right]_{1,2}$ = (4.8, 0.9) mK/K. Thus the relative contribution of the background to the total is $\frac{G_2}{G_1} \approx \frac{0.9}{4.8} = 19\%$ for the example data in

Supplementary Fig. 6. In other words, this analysis shows that the undesired parasitic heat transfers represent only 19% of the total measured heat transfer $G_1$ in this example, with the desired NFRHT comprising the substantial majority at 81%. The parasitic $G_2$ is subtracted before calculating the final $h_{rad}$ as reported in Fig. 2 of the main text, in this case a 19% correction. Similar calculations and corrections are repeated for all temperatures and for all test devices with their matched, co-fabricated reference devices. For the 50 and 20 nm thick devices, the analogous reference device measurements indicate that parasitic heat transfers are 23% and 35%, respectively, of the total heat transfer at 300 K.

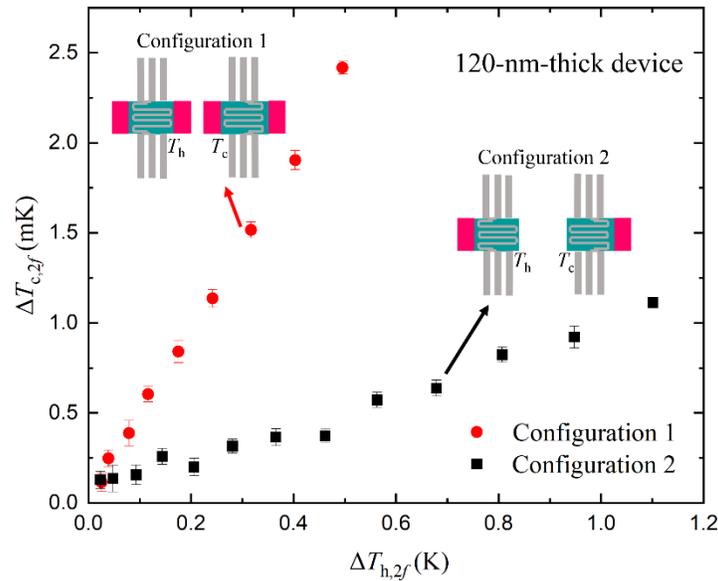

**Supplementary Fig. 6 | Measurement of parasitic heat transfer at 300 K on the 120 nm thick device.** Temperature rises in devices with and without SiC membranes are compared. On this plot the slopes of the Configuration 1 and Configuration 2 data, respectively, are proportional to the total and parasitic thermal conductances, denoted $G_1$ and $G_2$ in the text.

## 6. Convergence of DSGF simulations

We verified the convergence of the DSGF simulations with respect to the number of subvolumes used in the discretization as well as with respect to the membrane dimensions. The convergence analysis is shown hereafter solely for 120 nm thick device for brevity; the same convergence analysis was also performed on the 20 and 50 nm thick devices. As a motivating example, Supplementary Fig. 7 shows the spatial distribution of power density dissipated when the two membranes are discretized into a total of 13,312 nonuniform subvolumes. The simulated membrane length, $L$, and width, $w$, are both fixed at 1 μm in this simulation, which is significantly smaller than ($L = 9$ μm, $w = 30$ μm) in the experimental devices, so it is important to verify that the simulations are well converged for these smaller values of $L$ and $w$.

### *6.1. Number of subvolumes*

A nonuniform discretization scheme is used in the DSGF simulations in order to reduce the computational time[6]. Supplementary Fig. 7 shows that most of the heat is dissipated in the subvolumes adjacent to the gap spacing: for example, further analysis of the results reveals that 87% of the total power dissipated in the receiver is absorbed at $x$ locations within 500 nm of the gap, as indicated on the figure. Therefore, the finest subvolume sizes are assigned closest to the gap, with coarser subvolumes farther away from the gap. We then performed a grid refinement study of the geometry of Supplementary Fig. 7. Supplementary Fig. 8 shows the results for the total (i.e., spectrally integrated) radiative heat transfer coefficient, $h_{rad}$, at 300 K as a function of the number of subvolumes, for fixed $L = w = 1$ μm. Here, $h_{rad}$ is normalized by its maximum value, the number of subvolumes includes both discretized membranes, and a nonuniform discretization scheme is used. The configuration of Supplementary Fig. 7 corresponds to the second to last point of Supplementary Fig. 8. The percentage on top of each datapoint indicates the relative difference of that point's normalized $h_{rad}$ with respect to the previous point. These results show that $h_{rad}$ converges when the number of subvolumes is larger than 10,000, since the last three datapoints in Supplementary Fig. 8 have relative differences within 1%.

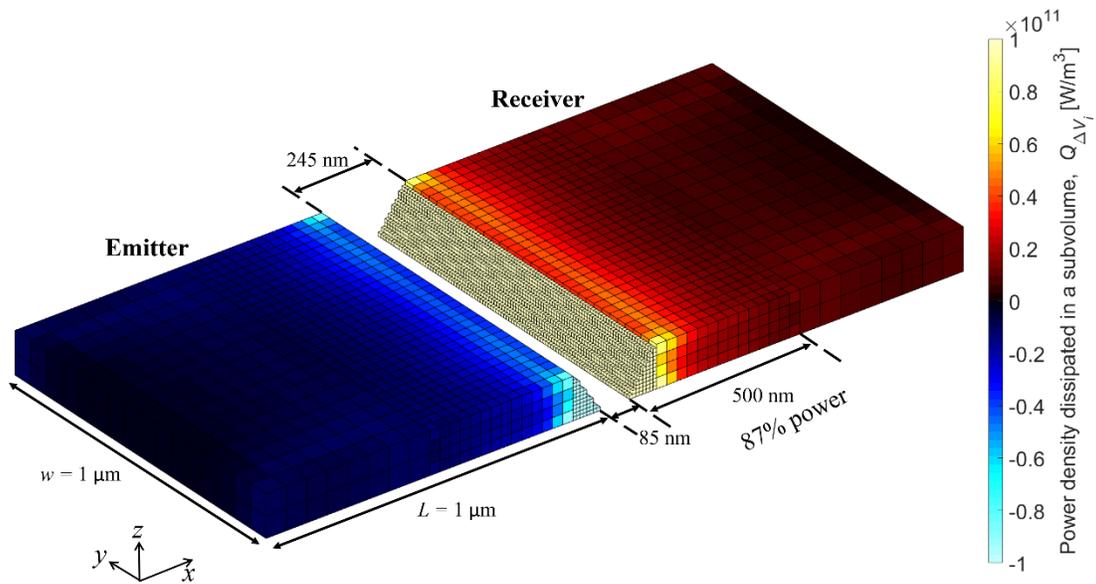

**Supplementary Fig. 7 | Example DSGF calculation of the spatial distribution of power density dissipated at 300 K between two 120 nm thick SiC membranes discretized into 13,312 nonuniform subvolumes.** Positive and negative values of the color bar represent heat gain and heat loss, respectively. The simulation considers the realistic nonuniform gap spacing between the emitter and receiver, with dimensions determined from the SEM image of Fig. 1e of the main text.

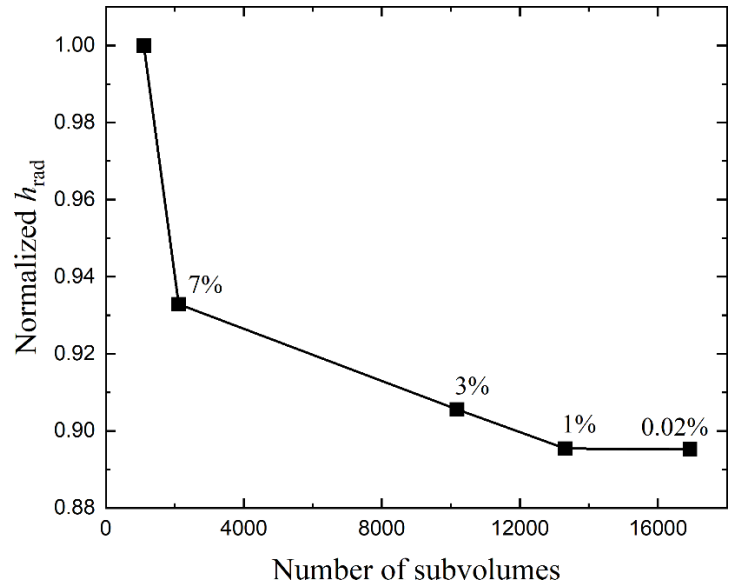

**Supplementary Fig. 8 | Normalized radiative heat transfer coefficient at 300 K between 120 nm thick SiC membranes as a function of the number of subvolumes.** The membrane length and width are fixed at 1 μm. The discretization is nonuniform, and the number of subvolumes includes both membranes. The heat transfer coefficient is normalized by its maximum value. The percentage listed on top of each data point indicates the relative difference of the normalized $h_{\text{rad}}$ with the previous point.

### *6.2. Membrane length*

The full 1 μm length of the 50 and 20 nm thick membranes is modeled in the DSGF simulations. For the 120 nm thick membranes, only a small portion of the 9 μm length needs to be modeled, since the power dissipated is very small in the subvolumes located far away from the gap (see Supplementary Fig. 7). Increasing the membrane length $L$ from 1 μm to 2 μm in the DSGF simulations was found to have a negligible impact on $h_{rad}$ (~3%, which is within the accuracy of the DSGF method[7]). As such, $L = 1$ μm is used for calculating the heat transfer coefficients for all membrane thicknesses.

### *6.3. Membrane width*

The DSGF simulations are challenging to perform for membrane widths significantly larger than 1 μm when using the discretization scheme shown in Supplementary Fig. 7 owing to large computational memory requirements[7]. Therefore, in order to keep the computational load to a manageable level, we analyze convergence with respect to the membrane width.

We use a base case of $t = 120$ nm and $L = 1$ μm, similar to Supplementary Fig. 7, and consider increasing $w$ of 1, 2, 4, and 8 μm. We discretize the membranes at a fineness corresponding to 2,112 nonuniform subvolumes per micron of $w$. Thus, the calculation for the pair of $w = 1$ μm membranes has 2,112 nonuniform subvolumes, while that for the pair of $w = 2$ μm membranes uses 4,224 nonuniform subvolumes, etc. While this level of discretization is somewhat coarse, it is still reasonable here for the purposes of this $w$ convergence study since, for the $w = 1$ μm case the total $h_{rad}$ calculated using 2,112 subvolumes differed by only 4% compared to the case of 13,312 subvolumes as already established in Supplementary Fig. 8. Furthermore the spectral $h_{rad}(\omega)$ results obtained with 2,112 and 13,312 subvolumes also are nearly identical (see Supplementary Fig. 9).

Supplementary Fig. 10 shows the total heat transfer coefficient, $h_{rad}$, at 300 K as function of the membrane width. As in Supplementary Fig. 8, $h_{rad}$ is normalized by its maximum value, and the relative differences are indicated on top of each datapoint. The total heat transfer coefficient is essentially insensitive to the membrane width, varying by less than 1.5% over this entire range of $w$ from 1 to 8 μm. The same conclusions apply for the 50 and 20 nm thick membranes (omitted for brevity). Therefore, the total heat transfer coefficients shown in Fig. 2 of the main text are calculated with membrane width of 1 μm for all devices. The corresponding numbers of nonuniform subvolumes are 13,312, 14,720, and 17,600 for the 120, 50, and 20 nm thick devices, respectively.

The spectral heat transfer coefficient converges for $w$ values slightly larger than 1 μm. Supplementary Fig. 11 shows the spectral $h_{\text{rad}}(\omega)$ for membrane widths of 1, 2, 4 and 8 μm. While the peak resonance frequencies are nearly identical for all membrane widths, for $w$ of 4 μm and below, the overall spectral shape shows some artifacts at lower frequency, i.e., in the spectral range 1.4 to $1.65 \times 10^{14}$ rad/s. We consider this $h_{\text{rad}}(\omega)$ adequately converged at $w$ = 8 μm and 16,896 subvolumes (using the discretization scheme based on 2,112 nonuniform subvolumes for $w$ = 1 μm). For the 50 and 20 nm thick membranes, similar calculations show that the spectral heat transfer coefficient converges for membrane widths of respectively 4 and 2 μm. Therefore, membrane widths of $w$ = 2, 4 and 8 μm have been used to calculate the spectral $h_{\text{rad}}(\omega)$ shown in Fig. 3a of the main text for the 20, 50, and 120 nm thick membranes, respectively.

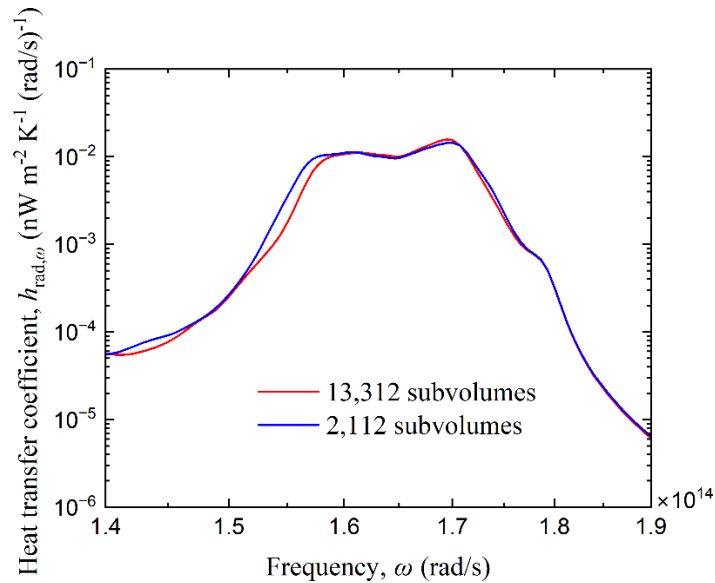

**Supplementary Fig. 9 | Spectral radiative heat transfer coefficient at 300 K between 120 nm thick SiC membranes for two different discretization schemes.** The membrane length and width are fixed at 1 μm. The discretization is nonuniform, and the number of subvolumes includes both discretized membranes. The 13,312 subvolume case is the same as shown in Supplementary Fig. 7.

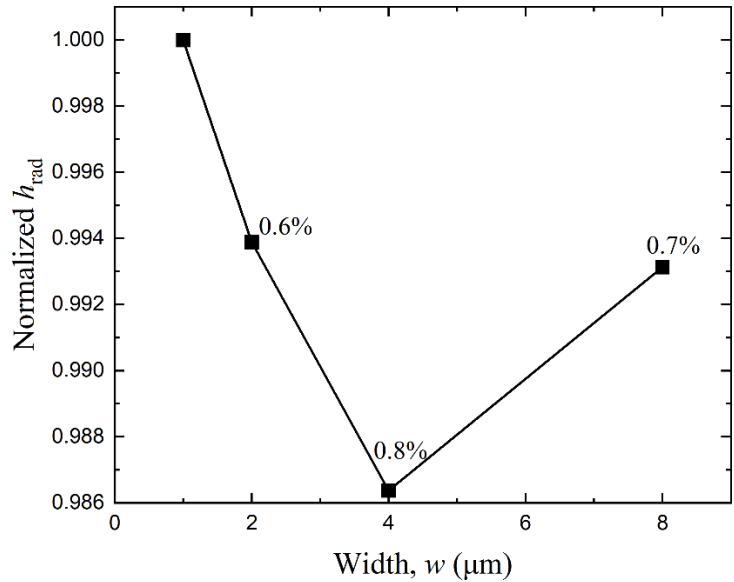

**Supplementary Fig. 10 | Normalized radiative heat transfer coefficient at 300 K between 120 nm thick SiC membranes as a function of their width.** The membrane length is 1 μm. The discretization scheme based on 2,112 nonuniform subvolumes per micron of width $w$ is used. The heat transfer coefficient is normalized by its maximum value. The percentage listed on top of each data point indicates the relative difference of its normalized $h_{rad}$ with the previous point.

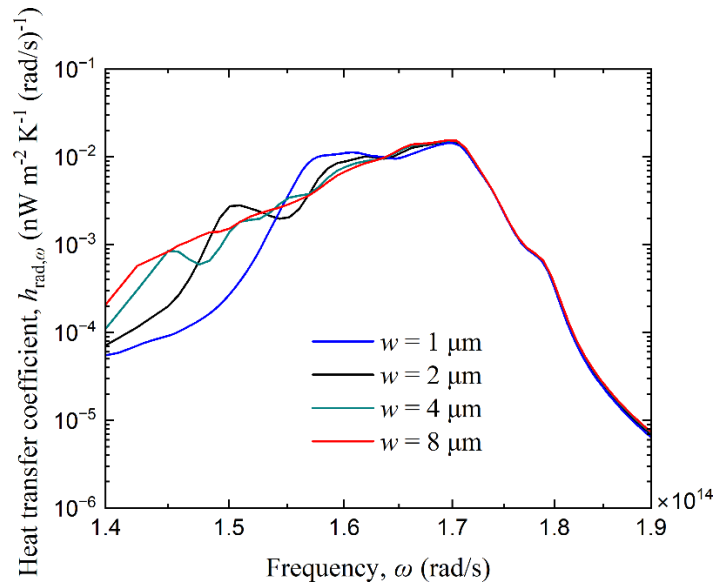

**Supplementary Fig. 11 | Spectral radiative heat transfer coefficient at 300 K between 120 nm thick SiC membranes as a function of their width.** The membrane length is 1 μm. The discretization scheme based on 2,112 nonuniform subvolumes per micron of width $w$ is used. The total integration of these curves corresponds to Supplementary Fig. 10.

## 7. Contribution from propagating waves to NFRHT between two SiC membranes

The contribution of propagating waves to NFRHT between SiC membranes is calculated using the DSGF method by considering only the far-field term in the free-space Green's function (see Methods). Supplementary Fig. 12 shows the propagating component of the heat transfer coefficient. The total radiative heat transfer, accounting for both propagating and evanescent waves, is also plotted for comparison. The contribution of propagating waves is negligibly small and accounts for less than 1% of the total heat transfer for all temperatures and membrane thicknesses. This indicates how the physics responsible for the heat transfer enhancement presented in this work is totally different than that described in Ref. 8 in which radiative heat transfer between subwavelength membranes is in the far-field regime and thus is solely mediated by propagating waves.

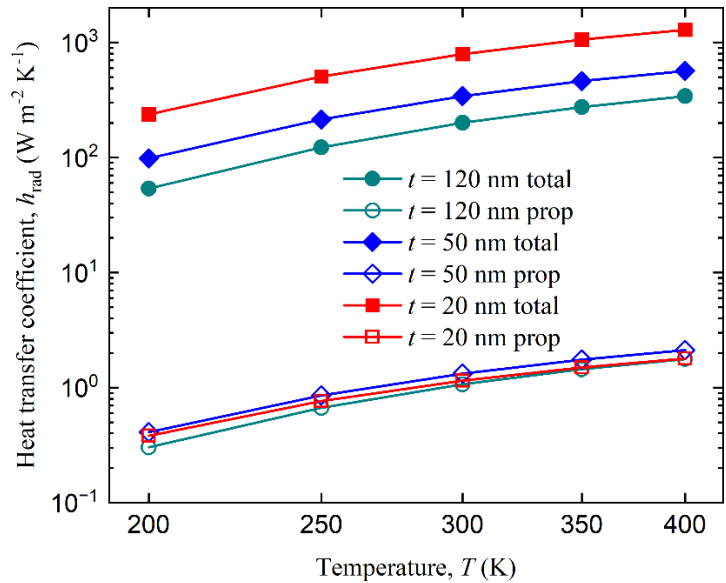

**Supplementary Fig. 12 | DSGF calculated contributions from propagating waves to NFRHT between two SiC membranes, using the *d* given in Supplementary Table 1.** The heat transfer coefficient due to propagating (prop) waves is plotted in hollow points. The total heat transfer coefficient, accounting for both propagating and evanescent waves, is also included for comparison with solid points (the results correspond to the case where the nominal values of the gap sizes are used and the tilts are not considered, or roughly the middle of the theoretical bands as shown in Fig. 2 of the main text). Propagating waves account for less than 1% of the total heat transfer for all temperatures and membrane thicknesses.

## 8. Additional thickness-dependent comparisons between experiments and DSGF simulations

Figure 2a of the main text presents the entirety of the experimental results along with the corresponding DSGF simulations, from which Fig. 2c isolates just the thickness dependence at 300 K. Supplementary Fig. 13 presents the analogous temperature "slices" like Fig. 2c but here for the four other temperatures, namely 200, 250, 350 and 400 K. The largest measured heat transfer coefficient is 1200 W/m$^2$K, observed for the $t$ = 20 nm device at 400 K. The largest enhancement is 5.7 if comparing to Reference Case 1 (NFRHT between infinite SiC surfaces) and 1400 if comparing to Reference Case 2 (blackbody limit), both also observed in the 20 nm device though now at 200 K.

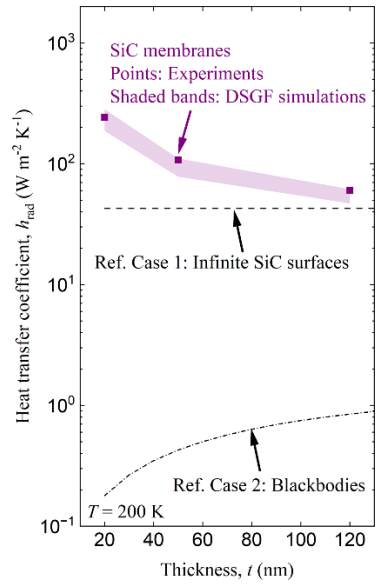

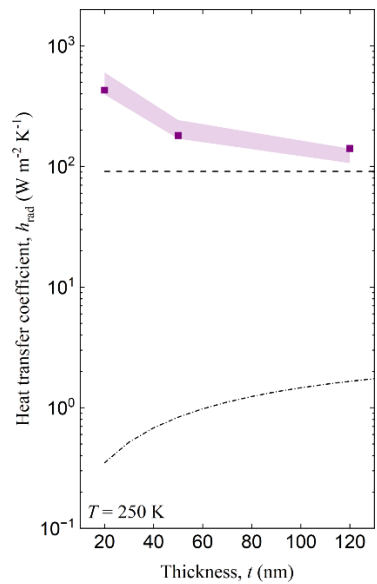

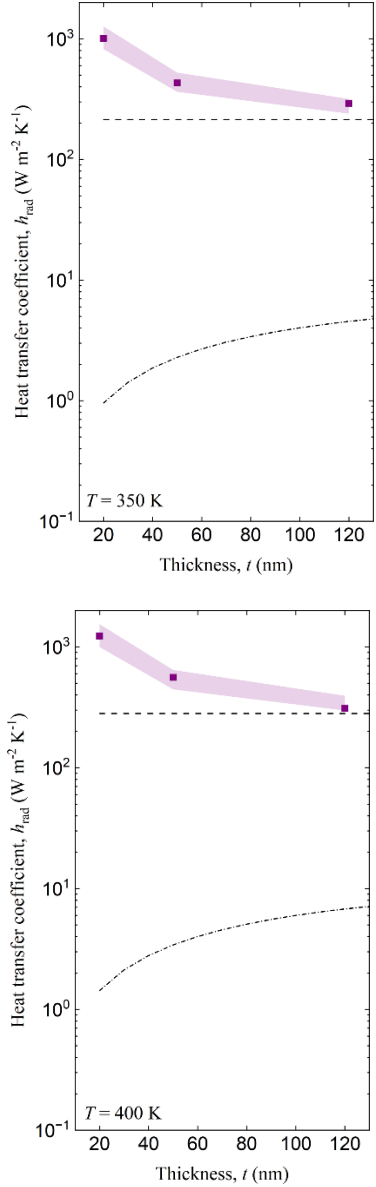

**Supplementary Fig. 13 | Measured and predicted temperature- and thickness-dependent radiative heat transfer coefficients.** These panels are constant-temperature slices for 200, 250, 350, and 400 K through Fig. 2a of the main text, completely analogous to the 300 K slice of Fig. 2c.

## 9. Verification of modal analysis

The accuracy of the modal analysis performed with COMSOL Multiphysics (Fig. 3b of main text) is verified here by comparison with a prior report by Berini[9] of the resonant modes supported by a solitary Ag membrane. We consider the modes in an Ag membrane with length $L = 1$ μm in $x$, infinite width $w \to \infty$ along $y$, and varying thickness $t$ from 0.02 to 0.20 μm along $z$ (see schematic in the inset of Supplementary Fig. 14). The Ag film is embedded in a lossless, infinite homogenous medium characterized by a dielectric function $\varepsilon_d = 4$. The frequency is fixed at $3 \times 10^{15}$ rad/s, and the dielectric function of Ag at this frequency is $\varepsilon_{Ag} = -19 - 0.53i$. We then use our COMSOL Multiphysics approach to calculate the real component of the wavevector in the $y$-direction, $k_y$, as a function of the film thickness. This yields four branches for each thickness since there are four fundamental modes: symmetric-symmetric ($ss$), symmetric-asymmetric ($sa$), asymmetric-asymmetric ($aa$), and asymmetric-symmetric ($as$)[9]. Supplementary Fig. 14 shows the resulting $k_y$ normalized by the vacuum wavevector, $k_0$, as a function of membrane thickness; the results from Ref. 9 are also plotted for comparison. The predictions based on COMSOL Multiphysics are in excellent agreement with those from Berini[9], which confirms the accuracy of the modal analysis. The modal analysis performed on the SiC membranes in the main text is similar to that for the Ag membranes. For the SiC membranes, the surrounding medium is vacuum ($\varepsilon_d = 1$), the membrane thickness is fixed, the frequency is varied, and all $k_y$ values at a specific frequency are calculated. The process is repeated for all membrane thicknesses. These results are shown in Fig. 3b of the main text.

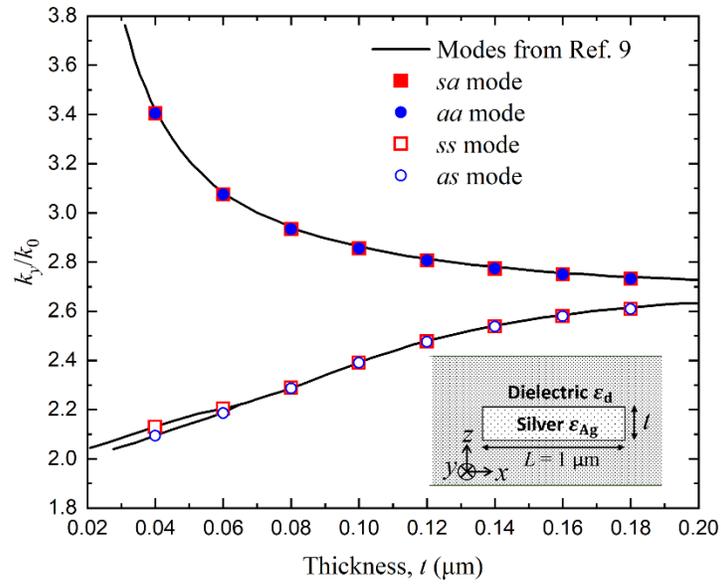

**Supplementary Fig. 14 | Modal analysis of an Ag membrane.** The real component of the wavevector in the $y$-direction, $k_y$, normalized by the vacuum wavevector, $k_0$, as a function of the Ag membrane thickness is calculated with COMSOL Multiphysics at a fixed frequency of $3 \times 10^{15}$ rad/s (points). The results from Ref. 9 are also plotted for comparison (solid line), demonstrating outstanding agreement.

## 10. Comparison of spectral radiative heat transfer coefficients calculated based on uniform and nonuniform gap spacings

The spectral radiative heat transfer coefficient shown in the main text (Fig. 3a) is calculated based on a uniform gap spacing of 100 nm in order to better explain the physics of NFRHT mediated by electromagnetic corner and edge modes. However, in the actual experiments, the gap spacings are nonuniform for the 50 and 120 nm thick devices, and this was already taken into account in the DSGF calculations of the total $h_{rad}$ in Fig. 2. Here we explore the effects of gap nonuniformity on the spectral $h_{rad}(\omega)$.

For the 50 and 120 nm films, Supplementary Fig. 15 compares the spectral $h_{rad}(\omega)$ calculated using the actual nonuniform gaps to that for the idealized uniform gaps of 100 nm as used in Fig. 3a. The magnitude of the heat transfer coefficient for nonuniform gap spacings is slightly smaller than that for uniform gap spacing owing to the larger overall separation distance between the membranes. The same trend is observed for uniform and nonuniform gap spacings: compared to the case of two infinite surfaces, the resonance broadens and redshifts as the membrane thickness decreases.

As for the case of uniform gaps, the redshifted resonances for nonuniform gap spacings observed in Supplementary Fig. 14 are mediated by electromagnetic corner and edge modes supported by subwavelength membranes (i.e., the resonance redshift is fully explained by the dispersion relation of Fig 3b). Yet, for a fixed membrane thickness, the resonance redshift is slightly more severe for nonuniform gap spacings due to larger average gaps. For example, the average nonuniform gap for the 120 nm thick device is $\bar{d} \approx 165$ nm as opposed to $d = 100$ nm used here for the uniform gap case. As explained in the main text, the largest contributing wavevector dominating NFRHT is estimated as $k_y \approx d^{-1}$. Therefore, for a larger gap, the largest contributing $k_y$ decreases, resulting in a slight resonance redshift as can be seen from the dispersion relation shown in Fig. 3b of the main text.

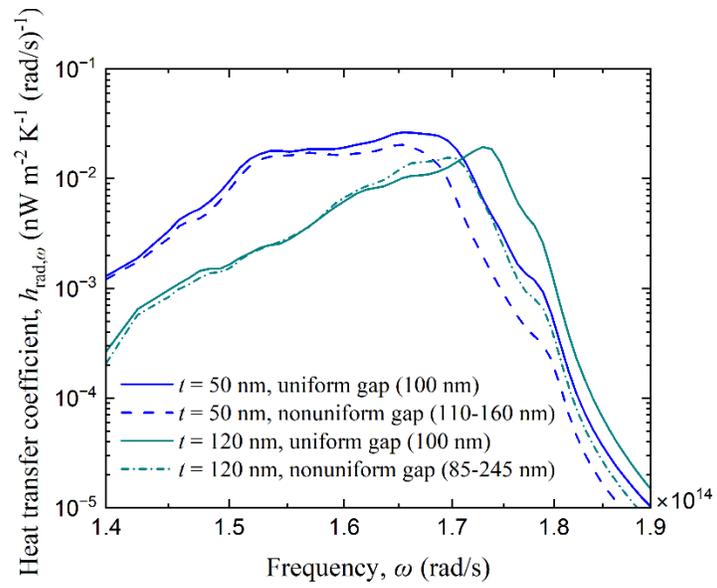

**Supplementary Fig. 15 | Spectral radiative heat transfer based on uniform and nonuniform gap spacings.** The spectral heat transfer coefficient is calculated at 300 K for the 50 and 120 nm thick devices.

## 11. Spatial distribution of electric and magnetic fields of the asymmetric-asymmetric (*aa*) mode

Supplementary Fig. 16 shows the spatial distribution of the electric and magnetic fields of the *aa* mode at a frequency of $1.61\times10^{14}$ rad/s in a 1-µm-long, infinitely wide, 20 nm thick SiC membrane obtained with COMSOL Multiphysics. This demonstrates that the *aa* mode is not purely TM-polarized, since all six field components exist (the same is true for the *as*, *sa*, and *ss* modes, not shown here). This is in marked contrast to the surface phonon-polariton modes allowed at an infinite planar SiC-vacuum interface, which are well known to be restricted exclusively to TM polarization.

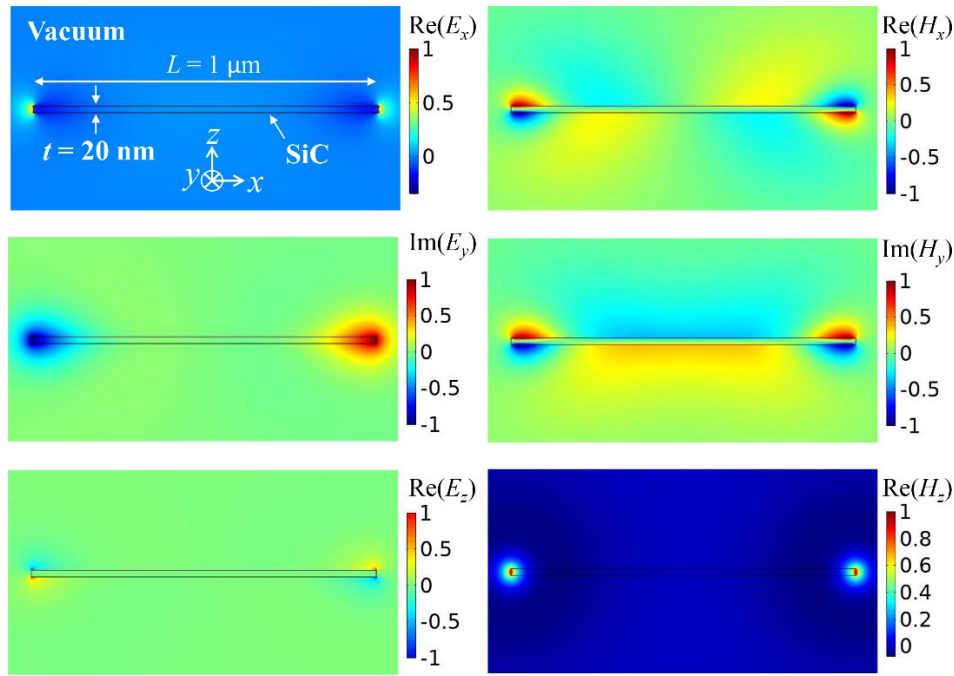

**Supplementary Fig. 16 | Spatial distribution of six field components of the *aa* mode at a frequency of $1.61\times10^{14}$ rad/s for a 1-µm-long, infinitely wide, 20 nm thick SiC membrane.** The field in each panel is normalized by its maximum value. As shown in the bottom-left panel, the *z*-component of the electric field, $E_z$, is asymmetric with respect to both the *z*- and *x*-axis, so following Ref. 9 this mode is labeled as asymmetric-asymmetric (*aa*).

## 12. Dielectric function of polycrystalline SiC

Supplementary Fig. 17 shows the real and imaginary components of the dielectric function of polycrystalline SiC as a function of frequency (see Methods). The imaginary part of the dielectric function increases as the frequency decreases in the Reststrahlen spectral band from the longitudinal optical phonon frequency, $\omega_{LO} = 1.801 \times 10^{14}$ rad/s, to the transverse optical phonon frequency, $\omega_{TO} = 1.486 \times 10^{14}$ rad/s. This implies that losses increase when the resonance of the heat transfer coefficient is redshifted (as long as $\omega$ remains above $\omega_{TO}$), thus resulting in spectral broadening. This helps explain the broadening seen in Fig. 3a.

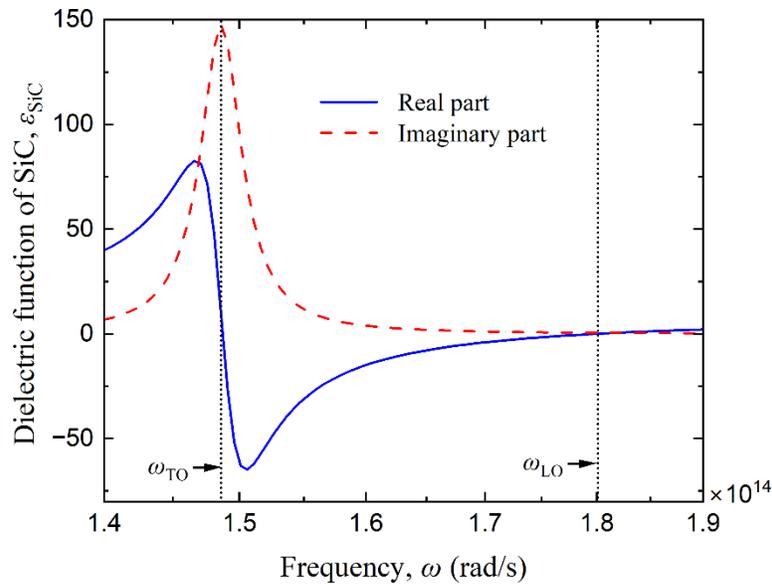

**Supplementary Fig. 17 | Frequency-dependent dielectric function of polycrystalline SiC used in the simulations.** The longitudinal optical phonon frequency $\omega_{LO}$ and the transverse optical phonon frequency $\omega_{TO}$ are also shown.